\documentclass[iicol, sn-vancouver]{sn-jnl}

\usepackage{booktabs}
\usepackage{array}
\usepackage{balance} 
\usepackage{lipsum}
\usepackage{multirow}
\usepackage{booktabs}
\usepackage{amsmath}
\usepackage{subfigure}
\usepackage{algorithm}  
\usepackage{algorithmicx}  
\usepackage{algpseudocode}  
\usepackage{amsmath}  
\usepackage{enumitem}
\usepackage{tabularx}
\usepackage[utf8]{inputenc}
\usepackage[english]{babel}
\usepackage{amsthm}
\usepackage{bm}
\usepackage{color}

\newcommand{\ie}{\emph{i.e., }}
\newcommand{\eg}{\emph{e.g., }}

\newcommand{\wrt}{\emph{w.r.t. }}
\newcommand{\cf}{\emph{cf. }}

\jyear{2021}%

\theoremstyle{thmstyleone}%
\theoremstyle{thmstyletwo}%

\theoremstyle{thmstylethree}%

\begin{document}

\title[Mitigating Spurious Correlations]{Mitigating Spurious Correlations for Self-supervised Recommendation}


\author[1]{\fnm{Xinyu} \sur{Lin}}
\equalcont{These two authors contributed equally to this work.}

\author[2]{\fnm{Yiyan} \sur{Xu}}
\equalcont{These two authors contributed equally to this work.}

\author[1]{\fnm{Wenjie} \sur{Wang}}

\author[2]{\fnm{Yang} \sur{Zhang}}

\author[2]{\fnm{Fuli} \sur{Feng}}

\affil[1]{\orgname{National University of Singapore}, \orgaddress{\country{Singapore}}}

\affil[2]{\orgname{University of Science and Technology of China}, \orgaddress{\country{China}}}


\abstract{
Recent years have witnessed the great success of self-supervised learning (SSL) in recommendation systems. However, SSL recommender models are likely to suffer from spurious correlations, leading to poor generalization. 
To mitigate spurious correlations, existing work usually pursues ID-based SSL recommendation or utilizes feature engineering to identify spurious features. Nevertheless, ID-based SSL approaches sacrifice the positive impact of invariant features, while feature engineering methods require high-cost human labeling. 
To address the problems, we aim to automatically mitigate the effect of spurious correlations. This objective requires to 1) automatically mask spurious features without supervision, and 2) block the negative effect transmission from spurious features to other features during SSL. 
To handle the two challenges, we propose an invariant feature learning framework, which first divides user-item interactions into multiple environments with distribution shifts and then learns a feature mask mechanism to capture invariant features across environments. Based on the mask mechanism, we can remove the spurious features for robust predictions and block the negative effect transmission via mask-guided feature augmentation. Extensive experiments on two datasets demonstrate the effectiveness of the proposed framework in mitigating spurious correlations and improving the generalization abilities of SSL models.
}




\maketitle

\section{Introduction}\label{sec:intro}

Self-supervised learning (SSL) approaches have recently become state-of-the-art (SOTA) for personalized recommendation \cite{wu2021self,yao2021self}. 
The core idea of SSL in recommendation is to learn better user and item representations via an additional self-discrimination task~\cite{zhou2020s3,wei2021contrastive}, which contrasts the augmentations over user-item features \cite{qian2022intent} or user-item interaction graphs \cite{wu2021self,xia2021self} to discover the correlation relationships among features and interactions~\cite{yao2021self}. 
Despite the great success, SSL-based recommender models are vulnerable to spurious correlations due to fitting the correlations from the input features to interactions. Because of the selection bias in the data collection process~\cite{pearl2009causality}, spurious correlations inevitably exist in the training data, where some spurious features show strong correlations with users' positive interactions (\eg clicks). 
\begin{figure}[t]
\setlength{\abovecaptionskip}{0cm}
\setlength{\belowcaptionskip}{0cm}
\centering
\includegraphics[scale=0.3]{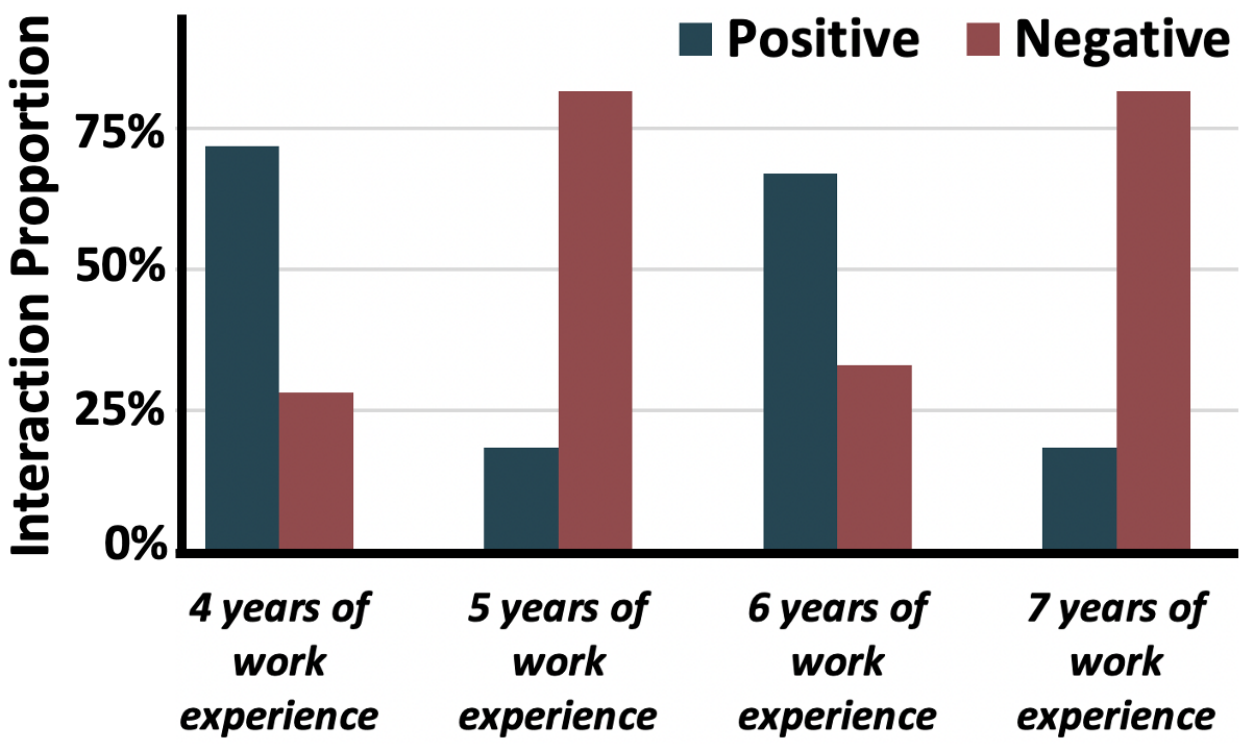}
\caption{An example of spurious correlations in job recommendation, where the positive interactions with the full-time jobs show strong correlations with the users' 4 and 6 years of experience.}
\label{fig:intro_example}
\end{figure}
As illustrated in Fig.~\ref{fig:intro_example}, users with 4 or 6 years of work experience are likely to have positive interactions with full-time jobs, while those with 5 or 7 years of experience are easy to have negative interactions. Such correlations between user experience and interactions are not reliable because users' preference for full-time jobs should not be significantly changed by only one year of experience gap. By the self-discrimination task, SSL models tend to capture these spurious correlations, resulting in poor generalization ability.

To alleviate the harmful effect of spurious correlations on SSL models, existing solutions mainly fall into three categories. Specifically,
\begin{itemize}[leftmargin=*]
    \item ID-based SSL methods \cite{wu2021self}, which only utilize IDs of users and items for collaborative filtering, and thus can avoid the harmful influence of some spurious features. However, the user and item features are still useful in the recommendation, especially for users with sparse interactions~\cite{yao2021self}. It is necessary to consider some invariant features that causally affect the interactions. For instance, accounting students usually prefer accountancy-related jobs.
    \item Feature engineering methods, which are able to identify a set of spurious features manually or using human-machine hybrid approaches \cite{DBLP:conf/hcomp/NushiKH18,DBLP:conf/icde/ChungKPTW19}. Thereafter, we can train the SSL recommender models by discarding the identified features. Nevertheless, feature engineering methods require extensive human-labeling work and thus are not applicable to large-scale recommendations with extensive user and item features. 
    \item Informative feature selection methods, which are capable of automatically recognizing the informative cross features and removing the redundant ones in the training process~\cite{cheng2020adaptive,autofis}. For instance, ~\cite{autofis} proposes a two-stage training strategy to identify informative feature interactions by a regularized optimizer and retrain the model after removing all the redundant features. Nevertheless, spurious features might be very informative for the interaction prediction in the training data, and thus degrade the generalization ability.
\end{itemize}

To solve the problems, we require the SSL models to automatically mitigate the effect of spurious correlations. 
In order to achieve this objective, there exist two essential challenges: 
\begin{itemize}[leftmargin=*]
    \item It is non-trivial to mask spurious features without supervision. SSL recommender models are expected to automatically identify the spurious features and drop them for robust predictions. As such, we should dig out the signals from the correlated data to guide the identification of spurious features.
    
    \item Blocking the effect transmission from spurious features to other features is of vital importance. SSL models usually maximize the features' mutual information via feature augmentation~\cite{Baruah2021data} (\eg correlated feature masking~\cite{yao2021self}) and contrastive learning, and thus the spurious features and other correlated features might have similar representations, transferring the detrimental effect of spurious features to other features. For example, users' experience in Fig.~\ref{fig:intro_example} might be correlated with the users' age, and SSL models are likely to learn similar representations for the experience and age via self-discrimination. 
\end{itemize}

To address the two challenges, we consider learning a feature mask mechanism from multiple environments to estimate the probabilities of spurious features and then adopt the mask mechanism to guide the feature augmentation in SSL models. Specifically, 1) we can cluster the interactions into multiple environments, where each environment has similar feature distributions, but the distributions shift between environments. The distribution shifts will guide the mask mechanism to capture invariant features across environments and exclude spurious features~\cite{scholkopf2021toward}. 
2) Besides, we can utilize the mask mechanism to drop the spurious features as the augmented sample and then maximize the mutual information between the invariant features in the augmented sample and all the input features in the factual sample, pushing SSL models to ignore the spurious features and cut off the negative effect transmission from spurious features to invariant features.

To this end, we propose an Invariant Feature Learning (IFL) framework for SSL recommender models to mitigate spurious correlations. In particular, IFL clusters the training interactions into multiple environments and leverages a masking mechanism with learnable parameters in $[0,1]$ to shield spurious correlations. To optimize the mask parameters, IFL adopts a variance loss to identify invariant features and achieve robust predictions across environments. As for the self-discrimination task, we drop the spurious features based on the mask parameters as the augmented sample, and then maximize the mutual information between the factual and augmented samples via contrastive loss, which pushes the SSL model to ignore spurious features. 
We instantiate IFL on a SOTA SSL model~\cite{yao2021self}, and extensive experiments on two real-world datasets validate the effectiveness of the proposed IFL in mitigating spurious correlations. 

In summary, our contributions are summarized as follows:

\begin{itemize}
    \item We point out the spurious correlations in SSL recommendation and consider learning invariant features from multiple environments.
    \item We propose a model-agnostic IFL framework, which leverages a feature mask mechanism and mask-guided contrastive learning to reduce spurious correlations for SSL models.
    \item Empirical results on two public datasets verify the superiority of our proposed IFL in masking spurious features and enhancing the generalization ability of SSL models.
\end{itemize}
\section{Method}\label{sec:method}
In this section, we first introduce the recommendation task, SSL recommendation, and spurious correlations in Section \ref{subsec:task_formulation}. 
And then, we present our IFL framework for SSL recommendation in Section \ref{subsec:invariant_feature_learning}, including feature mask learning and mask-guided contrastive learning. 

\subsection{Task Formulation}\label{subsec:task_formulation}
\subsubsection{Recommender Formulation}

The general idea of the recommendation task is to learn user preference from collected interactions between the users $\mathcal{U}$ and items $\mathcal{I}$, where each user $u\in\mathcal{U}$ has $N$ features such as ID, experience year, and country, denoted by $\bm{X}_{u}=\{\bm{x}_{u}^1,\bm{x}_{u}^2,\dots,\bm{x}_{u}^N\}$. 
Each entry in $\bm{X}_{u}$ is a one-hot vector indicating a specific feature value (\eg experience year$=$5). Likewise, an item $i$ has $M$ features $\bm{X}_{i}=\{\bm{x}_{i}^1,\bm{x}_{i}^2,\dots,\bm{x}_{i}^M\}$. 
Given a user-item interaction dataset $\mathcal{D}=\{(\bm{X}_u,\bm{X}_i, y_{ui})\}$ with $y_{ui}\in\{0,1\}$ indicating whether $u$ interacts with $i$, the recommender model aims to learn a function $f(u,i\vert\theta)$ to capture user preference. 
$\theta$ denotes the learnable parameters optimized over dataset $\mathcal{D}$ via the collaborative filtering (CF) loss, such as BPR loss~\cite{bpr}.




\subsubsection{Self-supervised Recommendation}

The SSL recommendation introduces an extra self-discrimination task to learn better user and item representations. The self-discrimination task includes two key steps: data augmentation and contrastive learning. 
SSL first augments factual samples by randomly dropping user and item features~\cite{yao2021self}, or conducting edge and node dropouts in the user-item interaction graph~\cite{wu2021self}. 
Then, based on the augmented samples, the positive and negative pairs are constructed for contrastive learning.

In the self-discrimination task, SSL models are essentially exploring the relationships between the user and item features, and the interactions~\cite{yao2021self}. As such, they are likely to fit spurious correlations from the spurious features to interactions. 

\subsubsection{Spurious correlations}
Spurious correlations broadly exist in the training dataset $\mathcal{D}$ due to the selection bias in data collection~\cite{pearl2009causality}. 
Intuitively, some spurious features do not causally affect the interactions but have a strong correlation with interactions purely because of the selection bias (see the example in Fig.~\ref{fig:intro_example}). 
By the normal recommender training over $\mathcal{D}$ and the additional self-discrimination task, SSL models will easily capture these shortcut correlations, suffering from poor generalization when the data distribution shifts. 
Moreover, the self-discrimination task via feature augmentation will maximize the mutual information between user and item features~\cite{yao2021self}, transferring the detrimental effect from spurious features to other invariant features. Such effect transmission in SSL models will further intensify the negative influence of spurious correlations. 

\subsection{Invariant Feature Learning}\label{subsec:invariant_feature_learning}
To mitigate the spurious correlations, we propose an IFL framework that can automatically identify spurious features via the feature mask mechanism and block the negative effect transmission by mask-guided contrastive learning. 

\begin{figure}[t]
\setlength{\abovecaptionskip}{0.1cm}
\setlength{\belowcaptionskip}{0.1cm}
\centering
\includegraphics[scale=0.35]{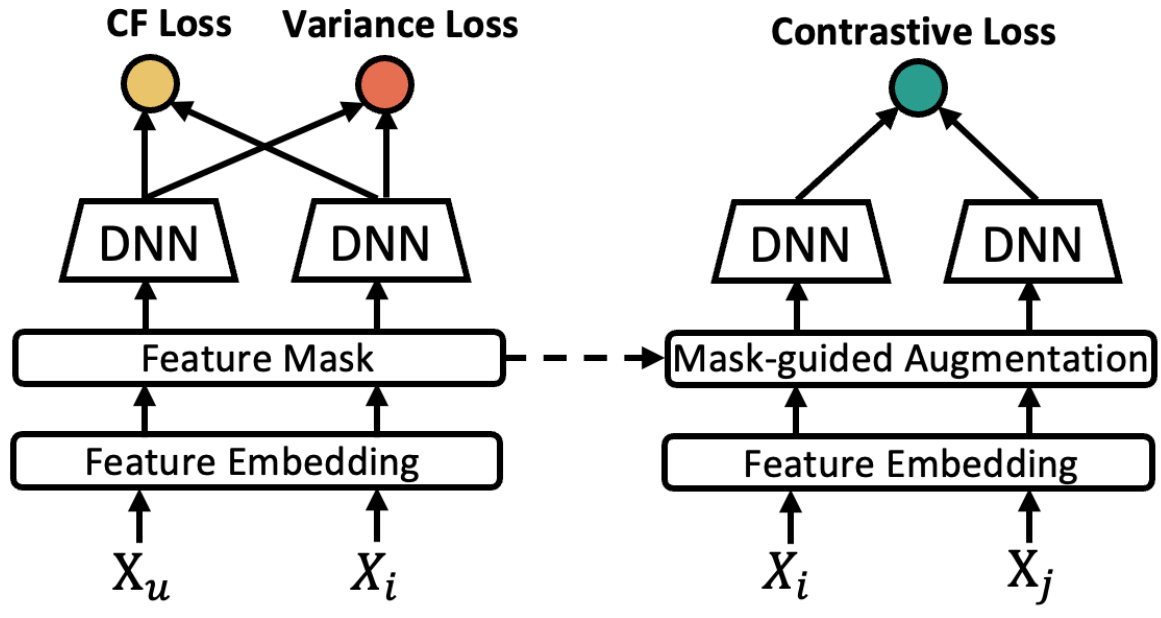}
\caption{Illustration of IFL framework. The two-tower encoders on the left are used for normal recommender training with the CF loss and variance loss. On the right side, we do mask-guided feature augmentation over the user and item features, \ie drop the spurious features and then conduct contrastive learning. The augmentations on users and items are the same and we only show that on items to save space.}
\label{fig:framework}
\end{figure}


The overall IFL framework is demonstrated in Fig. \ref{fig:framework}.
Given a pair of user and item features $(\bm{X}_u$,$\bm{X}_i)$, we first look up their embeddings via the feature embedding layer, and then utilize the feature mask mechanism to mask the embeddings of spurious features for predictions. We utilize the CF loss to optimize the feature embeddings while adopting a variance loss to supervise the learning of the mask mechanism. Regarding self-discrimination, we feed the user and item features into the mask-guided augmentation layer and conduct contrastive learning over the factual and augmented samples. 


\subsubsection{Feature Mask Learning}\label{subsubsec:feature_mask_learning}
%
In order to identify spurious features and remove their harmful influence, we introduce a feature mask mechanism. 

Specifically, we define two feature masks $\bm{m}_u \in \mathbb{R}^N$, and $\bm{m}_i \in \mathbb{R}^M$, which are shared with all users and items, respectively. $\bm{m}_u$ and $\bm{m}_i$ with the range $[0,1]$, denoting the probability of being invariant features, and thus a feature with a smaller mask value is more likely to be a spurious feature. 
To estimate the two masks, we draw $\bm{m}_u$, and $\bm{m}_i$ from the clipped Gaussian distributions~\cite{yamada2020feature} parameterized by $\bm{\gamma}_1 \in \mathbb{R}^N$ and $\bm{\gamma}_2 \in \mathbb{R}^M$, respectively. Formally, for each $m_u[k] \in \bm{m}_u$ and $m_i[j] \in \bm{m}_i$, we have
\begin{equation}
\label{eqn:clip}
\left\{
\begin{aligned}
&\bm{m}_u[k] = \text{min}(\text{max}(\bm{\gamma}_1[k] + \epsilon, 0),1), \\
&\bm{m}_i[j] = \text{min}(\text{max}(\bm{\gamma}_2[j] + \epsilon, 0),1), \\
\end{aligned}
\right .
\end{equation}
where the noise $\epsilon$ is drawn from $\mathcal{N}\left(0, \sigma_{\epsilon}^2\right)$. 
To ensure the two masks well represent the probabilities, we clip the values of $\bm{m}_u$ and $\bm{m}_i$ into $[0,1]$.
Thereafter, we apply the masks to shield the spurious features before feeding the feature embeddings into the Deep Neural Network (DNN) encoders. In detail, we represent the user and item features via the embeddings $\bm{X}_u\footnote{To keep notation brevity, we use $\bm{X}_u$ to represent both users' input features and their embeddings. It is similar for $\bm{X}_i$.}\in \mathbb{R}^{N\times D}$ and $\bm{X}_i \in \mathbb{R}^{M\times D}$, where $D$ is the embedding dimension. 
For each $\bm{X}_u[:,k] \in \bm{X}_u$ and $\bm{X}_i[:,k] \in \bm{X}_i$, we have
\begin{equation}
\label{eqn:dropped_features}
\left\{
\begin{aligned}
&\bm{X}'_u[:,k]=\bm{X}_u[:,k] \odot \bm{m}_u, \\
&\bm{X}'_i[:,k]=\bm{X}_i[:,k] \odot \bm{m}_i, \\
\end{aligned}
\right .
\end{equation}
where $\odot$ is the element-wise multiplication. 
As such, the feature embeddings are masked via $\bm{m}_u$ and $\bm{m}_i$.
Next, we concatenate $N$ feature embeddings in $X'_u$ into a vector and then feed it into the DNN encoders to obtain the user representation $\bm{z}_u$ and item representation $\bm{z}_i$.


\begin{figure}[t]
\setlength{\abovecaptionskip}{0cm}
\setlength{\belowcaptionskip}{0cm}
\centering
\hspace{-0.2in}
\includegraphics[scale=0.25]{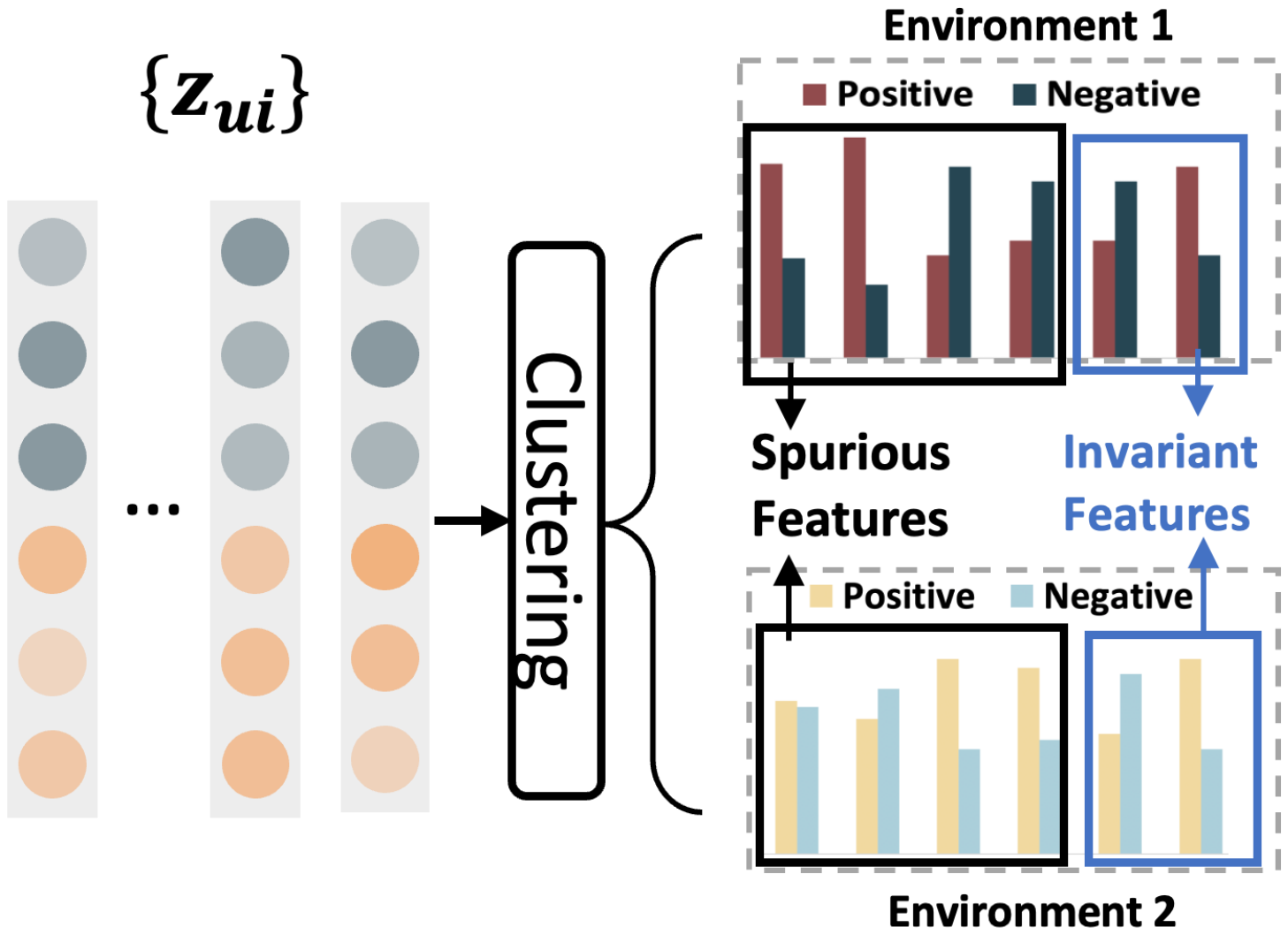}
\caption{Illustration of the environment division in feature mask learning. The interaction representations $\bm{z}_{ui}$ are clustered into $C=2$ environments. The distributions of spurious features shift across two environments while those of invariant features are stable.
}
\label{fig:split_environment}
\end{figure}

\vspace{5pt}
\noindent$\bullet\quad$\textbf{Environment Division.} 
The key challenge of automatically recognizing spurious features is to extract the supervision signals from the correlated data. 
To address this challenge, we cluster the interactions into multiple environments with distribution shifts and utilize the shifts to discover invariant features. 
Specifically, we obtain the representation of each user-item pair $\bm{z}_{ui}$ by concatenating the user and item representations:
\begin{equation}
\label{eqn:interaction_rep}
\begin{aligned}
\bm{z}_{ui}=\text{concat}(\bm{z}_u,\bm{z}_i).
\end{aligned}
\end{equation}
Thereafter, we adopt K-means to cluster the interaction representations $\bm{z}_{ui}$ into $C$ environments:
\begin{equation}
\label{eqn:clustering}\notag
\begin{aligned}
\{\bm{z}_{ui}^c\vert c\in \{1,2,\dots,C\}\} \leftarrow \text{K-means}(\{\bm{z}_{ui}\}),
\end{aligned}
\end{equation}
where $\bm{z}_{ui}^c$ denotes the interaction representation that belongs to environment $c$. 
By clustering, similar features will be divided into the same environment while the feature distribution shifts across environments. 
The interactions with the same spurious features are similar and thus are easy to be clustered into the same environments. 
As illustrated in Fig.~\ref{fig:split_environment}, the spurious features show different distributions across environments, and only invariant features have consistent distributions. 
As such, pursuing robust predictions across environments will push the mask mechanism to discover the invariant features and exclude spurious features.  

\vspace{5pt}
\noindent$\bullet\quad$\textbf{CF and Variance Loss.} 
To push the mask mechanism to exclude spurious features, we incorporate a variance loss. 
Before the variance loss, we first detail the CF loss for the normal recommender training. Following~\cite{yao2021self}, we adopt the batch-softmax loss for a batch of interactions, \ie
\begin{equation}
\label{eqn:CF_loss}
\begin{aligned}
\mathcal{L}_{\text{CF}}=-\frac{1}{B}\sum_{k=1}^{B}\log\frac{\exp(s(\bm{z}_{u_k},\bm{z}_{i_k})/\tau)}{\sum_{j=1}^{B}\exp(s(\bm{z}_{u_k},\bm{z}_{i_j})/\tau)},
\end{aligned}
\end{equation}
where $B$ is the batch number, $\tau$ refers to a temperature hyper-parameter, and $s(\cdot)$ denotes the cosine similarity function. 

As to the variance loss, we separately calculate $\mathcal{L}_{\text{CF}}$ within each environment, \ie $\mathcal{L}_c$ with $c$ varying from $1$ to $C$. Formally,
\begin{equation}
\label{eqn:L_c}
\begin{aligned}
\mathcal{L}_c=-\frac{1}{B_c}\sum_{k=1}^{B_c}\log\frac{\exp(s(\bm{z}^c_{u_k},\bm{z}^c_{i_k})/\tau)}{\sum_{j=1}^{B_c}\exp(s(\bm{z}^c_{u_k},\bm{z}^c_{i_j})/\tau)},
\end{aligned}
\end{equation}
where $\bm{z}^c_{u_k}$, $\bm{z}^c_{i_k}$ represent the $k$-th user and item representation of environment $c$, and $B_c$ is the number of interactions that belong to environment $c$.
And then, we regulate the mask mechanism by minimizing the gradient variance $\mathcal{L}_v=\mathcal{L}_{v_u}+\mathcal{L}_{v_i}$, where $\mathcal{L}_{v_u}$ and $\mathcal{L}_{v_i}$ are as follows:
\begin{equation}
\label{eqn:variance_loss}
\left \{
\begin{aligned}
&\mathcal{L}_{v_u}=\sum_{c=1}^C\left\| \nabla_{\bm{m}_u} \mathcal{L}_c - \nabla_{\bm{m}_u} \mathcal{L}_{\text{CF}}\right\|^2,\\
&\mathcal{L}_{v_i}=\sum_{c=1}^C\left\| \nabla_{\bm{m}_i} \mathcal{L}_c - \nabla_{\bm{m}_i} \mathcal{L}_{\text{CF}}\right\|^2, \\
\end{aligned}
\right .
\end{equation}
where $\nabla_{\bm{m}_u} \mathcal{L}_c$ is the gradients \wrt the mask $\bm{m}_u$ in environment $c$ and $\nabla_{\bm{m}_u} \mathcal{L}_{\text{CF}}$ denotes the average gradients \wrt $\bm{m}_u$ across $C$ environments. $\mathcal{L}_{v_u}$ thus reflects the gradient variance in $C$ environments. Here, the utility of gradient variance follows the idea of invariant learning in \cite{HRM}, where the gradient variance is simplified from the variance penalty regularizer proposed in \cite{MIPkoyama} for feature selection scenarios. Minimizing the gradient variance will regulate $\bm{m}_u$ to have similar gradients and close performance in multiple environments~\cite{IRM}. The gradient loss over $\mathcal{L}_{v_i}$ is similar to $\mathcal{L}_{v_u}$ for the regularization of $\bm{m}_i$.
As such, optimizing the variance loss avoids the situation that most environments are well predicted via capturing spurious correlations, while few environments have inferior performance because of the spurious correlations. 


\subsubsection{Mask-guided Contrastive Learning}\label{subsubsec:mask-guided_contrastive_learning}
By incorporating the feature mask mechanism into the self-discrimination task, we require the model to ignore the spurious features and cut off the negative effect transmission from spurious features to invariant features.

\begin{figure}[t]
\setlength{\abovecaptionskip}{0cm}
\setlength{\belowcaptionskip}{0cm}
\centering
\includegraphics[scale=0.25]{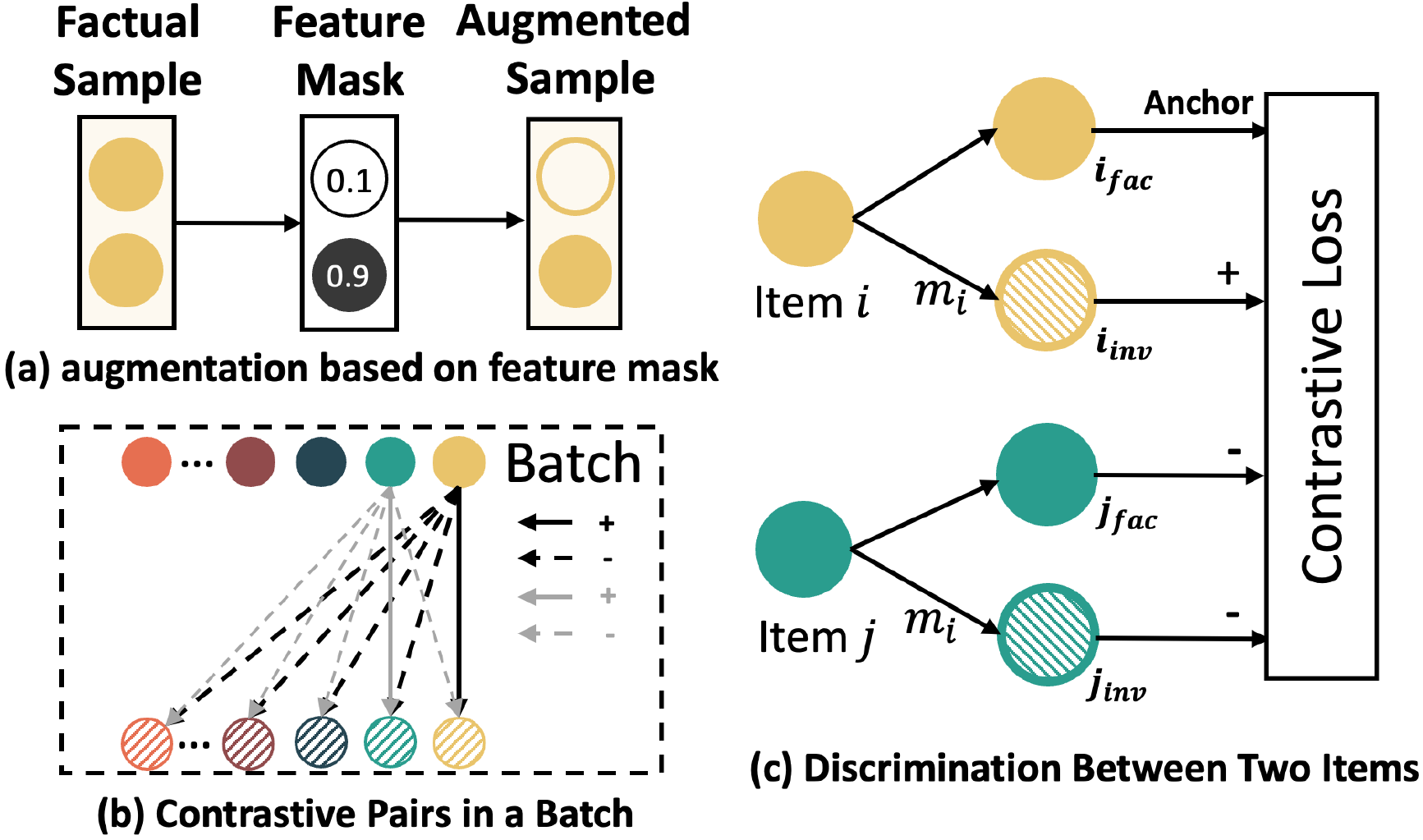}
\caption{Illustration of mask-guided contrastive learning in IFL. (a) shows the augmentation by dropping spurious features, (b) demonstrates the contrastive pairs in a sample batch, and (c) presents the contrastive loss between two items. Only two samples' contrastive pairs are shown in (b) for neatness.}
\label{fig:ours_contrastive}
\end{figure}

\vspace{5pt}
\noindent$\bullet\quad$\textbf{Data Augmentation.} 
Instead of random feature dropout for augmentation, we consider a dropping probability to drop the spurious features. The dropping probability of each feature is proportional to its probability of being a spurious feature in the mask mechanism. As shown in Fig.~\ref{fig:ours_contrastive}, we then utilize contrastive learning to maximize the mutual information between the factual sample and augmented samples. 
\vspace{5pt}

\noindent$\bullet\quad$\textbf{Contrastive Loss.} 
Fig.~\ref{fig:ours_contrastive}(c) illustrates the contrastive loss in a batch of samples. Formally,
\begin{equation}
\label{eqn:ours_contrastive_loss}
\begin{aligned}
\mathcal{L}_{\text{ssl}}=-\frac{1}{B}\sum_{k=1}^B\log\frac{\exp(s(\bm{z}_{{k_\text{fac}}},\bm{z}_{{k_\text{inv}}})/\tau)}{\sum_{j=1}^B\exp(s(\bm{z}_{{k_{\text{fac}}}},\bm{z}_{{j_\text{inv}}})/\tau)},
\end{aligned}
\end{equation}
where $\bm{z}_{k_{fac}}$ stands for the representation of a factual sample with all input features, and $\bm{z}_{k_{inv}}$ is for the augmented sample that drops spurious features. By considering that $\bm{z}_{k_{fac}}$ and $\bm{z}_{k_{inv}}$ are positive pairs, the SSL models will ignore the spurious features recognized by the mask mechanism, blocking the effect transmission from spurious features to other invariant features. 

\vspace{5pt}
\noindent$\bullet\quad$\textbf{Optimization.} In our proposed IFL, the overall objective function is formulated as:
\begin{equation}
\label{eqn:objective_function}
\begin{aligned}
\mathcal{L}=\mathcal{L}_{\text{CF}}+\alpha\mathcal{L}_{\text{ssl}}+\beta\mathcal{L}_v+ \lambda{\|\theta\|}^2,
\end{aligned}
\end{equation}
where $\theta$ denotes the model's learnable parameters, and $\|\theta\|^2$ is the L2 norm regularization term to avoid over-fitting. Besides, $\alpha$, $\beta$, and $\theta$ are the hyper-parameters to adjust the effect of the contrastive loss, variance loss, and regularization. Finally, we utilize this overall loss to optimize the parameters by gradient descent. 
\section{Experiment}\label{sec:exp}
We conduct extensive experiments to answer the following research questions:
\begin{itemize}[leftmargin=*]
    \item \textbf{RQ1:} Can the IFL framework outperform baselines on the recommendation performance when spurious features exist?
    
    \item \textbf{RQ2:} 
    How do the components of IFL affect its effectiveness?
    
    \item \textbf{RQ3:} How do the hyper-parameters affect the validity of the proposed method?
    
    
\end{itemize}

\subsection{Experimental Settings}

\vspace{5pt}
\noindent\textbf{Datasets.} We conduct experiments on two real-world datasets, Meituan\footnote{\url{https://www.biendata.xyz/competition/smp2021_2/.}} and  XING\footnote{\url{http://www.recsyschallenge.com/2017/.}}. The statistics of them are shown in Table~1.

\begin{itemize}[leftmargin=*]
    \item Meituan is a food recommendation dataset with rich user consumption of food. For each sample, we keep the following important attributes: user ID, user income, item ID, and item price.

    
    \item XING is a job recommendation dataset,  including several types of interactions and abundant features of users and jobs. For simplicity, we delete some unimportant features. We merge three types of interactions (types of $1-3$) to reflect user interests, making a sample negative only if the three types of interaction are all negative.
    
\end{itemize}

\begin{figure}[tb]
    \setlength{\abovecaptionskip}{0cm}
    \setlength{\belowcaptionskip}{-0.2cm}
    \centering
    \includegraphics[scale=0.3]{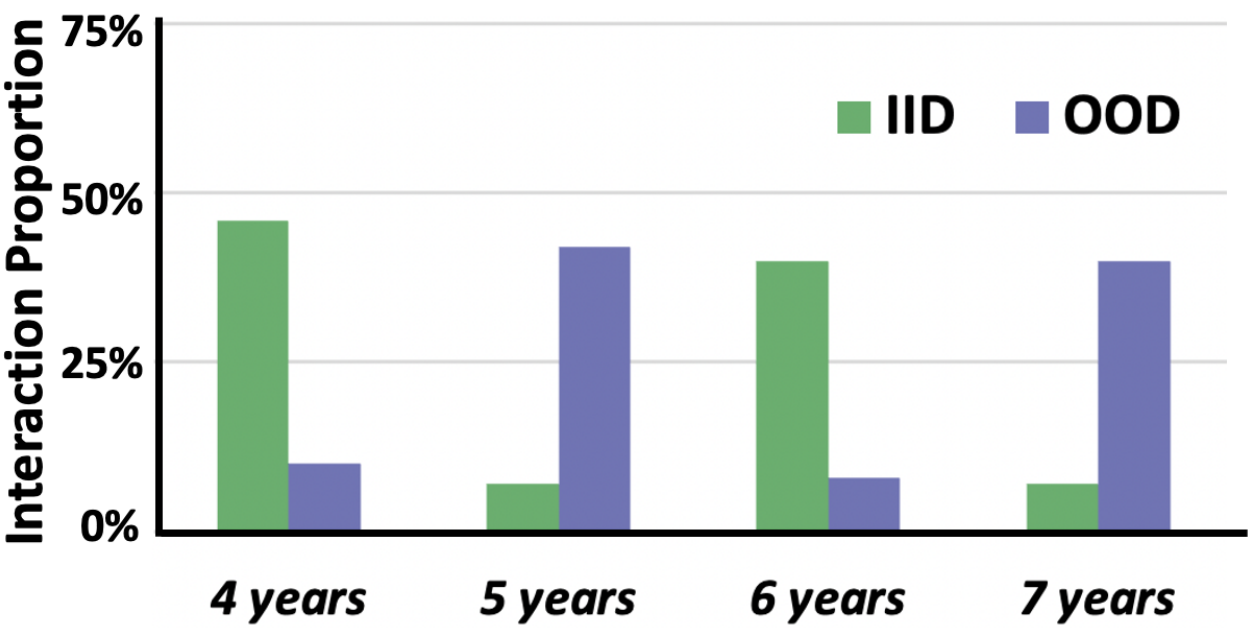}
    \caption{The conditional distribution of the interaction given spurious features for the IID and OOD sets of XING, where the spurious feature is the user work experience and the interaction is about interacting with the full-time job.}
    \label{fig:distribution}
\end{figure}

We apply the 5-core setting~\cite{ngcf} to filter the datasets and randomly split them into training, validation, and testing sets with the ratio of 8:1:1, where the testing set is used for the IID testing setting, \ie the setting that the training and testing have the independent and identical distribution. Besides, to better evaluate the models' ability to deal with spurious correlations, we also construct an OOD testing set, in which the spurious correlations are manually controlled.
For example, we discover the work experience as the spurious feature for dataset XING. Then we forcibly filter out some interactions such that the OOD set and the training set have significantly different conditional distributions of the interaction given this spurious feature, as shown in Fig.~\ref{fig:distribution}.


\begin{table}[t]
\label{table:statistics_of_dataset}
\setlength{\abovecaptionskip}{0cm}
\setlength{\belowcaptionskip}{0cm}
\setlength{\tabcolsep}{1.5mm}{\caption{The statistics of the two datasets. Here, 'Int.' denotes 'Interactions'.}}
\begin{center}
\setlength{\tabcolsep}{2mm}{
\resizebox{0.48\textwidth}{!}{
\begin{tabular}{lcccc}
\hline
 \textbf{Dataset} & \multicolumn{1}{l}{\textbf{\#Users}} & \multicolumn{1}{l}{\textbf{\#Items}} & \multicolumn{1}{l}{\textbf{\#Int.}} & \multicolumn{1}{l}{\textbf{Density}} \\ \hline
{Meituan} & 2113 & 7138 & 13429 & 0.00089 \\
{XING} & 22668 & 8756 & 120976 & 0.00061 \\ \hline
\end{tabular}
}
}
\end{center}
\end{table}

\begin{table*}[t]
\setlength{\abovecaptionskip}{0cm}
\setlength{\belowcaptionskip}{0cm}
\setlength{\tabcolsep}{1.5mm}{\caption{Performance comparisons between baselines and IFL on Meituan and XING. The best results are highlighted in bold and the sub-optimal results are underlined.}}
\label{table:overall}
\begin{center}
\setlength{\tabcolsep}{2mm}{
\resizebox{\textwidth}{!}{
\begin{tabular}{lcccccccc}
\hline
\multicolumn{9}{c}{\textbf{Meituan}} \\ \hline
\multicolumn{1}{l|}{}                    & \multicolumn{4}{c|}{\textbf{OOD testing}}                                                            & \multicolumn{4}{c}{\textbf{IID testing}}                                        \\
\multicolumn{1}{l|}{\textbf{Model}}      & \textbf{Recall@50} & \textbf{Recall@100} & \textbf{NDCG@50} & \multicolumn{1}{c|}{\textbf{NDCG@100}} & \textbf{Recall@50} & \textbf{Recall@100} & \textbf{NDCG@50} & \textbf{NDCG@100} \\ \hline
\multicolumn{1}{l|}{\textbf{FM}}         & 0.0144             & 0.0231              & 0.0043           & \multicolumn{1}{c|}{0.0058}            & 0.0656             & 0.0789              & 0.0280            & 0.0302            \\
\multicolumn{1}{l|}{\textbf{NFM}}        & 0.0139             & 0.0252              & 0.0039           & \multicolumn{1}{c|}{0.0057}            & 0.0597             & 0.0761              & 0.0260            & 0.0287            \\
\multicolumn{1}{l|}{\textbf{DeepFM}}     & 0.0159             & \underline{0.0257}        & 0.0041           & \multicolumn{1}{c|}{0.0057}            & 0.0618             & 0.0763              & 0.0257           & 0.0281            \\
\multicolumn{1}{l|}{{\textbf{AutoDeepFM}}}   & {0.0109}           & {0.0227}              & {0.0031}           & \multicolumn{1}{c|}{{0.0050}}             & {0.0646}            & {0.0794}             & {0.0266}          & {0.0290}            \\ 
\multicolumn{1}{l|}{{\textbf{AutoFIS+CFM}}}   & {0.0128}           & {0.0194}              & {0.0037}           & \multicolumn{1}{c|}{{0.0048}}             & {0.0590}            & {0.0824}             & {0.0205}          & {0.0243}            \\
\multicolumn{1}{l|}{{\textbf{SGL+Attr-A}}}   & {0.0166}          & {0.0251}          & {0.0056}         & \multicolumn{1}{c|}{{0.007}}          & {0.0792}        & {0.0929}        & {0.0354}        & {0.0377}         \\
\multicolumn{1}{l|}{{\textbf{SGL+Attr-B}}}   & {0.0047}          & {0.0118}          & {0.0011}         & \multicolumn{1}{c|}{{0.0022}}          & {0.0374}        & {0.0509}        & {0.0137}        & {0.0159}         \\
\multicolumn{1}{l|}{\textbf{SGL}}        & \underline{0.0180}        & 0.0246              & \underline{0.0061}     & \multicolumn{1}{c|}{\underline{0.0072}}      & \underline{0.0896}       & \underline{0.1047}        & \underline{0.0357}     & \underline{0.0382}      \\
\multicolumn{1}{l|}{\textbf{SSL-DNN}}   & 0.0123             & 0.0232              & 0.0042           & \multicolumn{1}{c|}{0.0060}             & 0.0763             & 0.1015              & 0.0244           & 0.0285            \\ 
\multicolumn{1}{l|}{\textbf{IFL}}       & \textbf{0.0227}    & \textbf{0.0345}     & \textbf{0.0064}  & \multicolumn{1}{c|}{\textbf{0.0083}}   & \textbf{0.1013}    & \textbf{0.1159}     & \textbf{0.0413}  & \textbf{0.0436}   \\ \hline 
\multicolumn{9}{c}{\textbf{XING}} \\ \hline
\multicolumn{1}{c|}{}                  & \multicolumn{4}{c|}{\textbf{OOD testing}}                                                         & \multicolumn{4}{c}{\textbf{IID testing}}                               \\
\multicolumn{1}{c|}{}                  & \textbf{Recall@10}       & \textbf{Recall@20}      & \textbf{NDCG@10}        & \multicolumn{1}{c|}{\textbf{NDCG@20}}         & \textbf{Recall@10}     & \textbf{Recall@20}     & \textbf{NDCG@10}       & \textbf{NDCG@20}       \\ \hline
\multicolumn{1}{l|}{\textbf{FM}}       & 0.5402          & 0.7056         & 0.3003         & \multicolumn{1}{c|}{0.3505}          & 0.6163        & 0.6605        & 0.3998        & 0.4117        \\
\multicolumn{1}{l|}{\textbf{NFM}}      & 0.5425          & 0.7076         & 0.3069         & \multicolumn{1}{c|}{0.3569}          & 0.6202        & 0.6672        & 0.4033        & 0.4160         \\
\multicolumn{1}{l|}{\textbf{DeepFM}}   & 0.5431          & 0.7110          & 0.3066         & \multicolumn{1}{c|}{0.3575}          & 0.6223        & 0.6678        & 0.4047        & 0.4170         \\
\multicolumn{1}{l|}{{\textbf{AutoDeepFM}}}   & {0.5312}           & {0.6977}              & {0.3006}           & \multicolumn{1}{c|}{{0.3510}}             & {0.6182}            & {0.6572}             & {0.4014}          & {0.4120}            \\ 
\multicolumn{1}{l|}{{\textbf{AutoFIS+CFM}}}   & {0.5374}           & {0.7062}              & {0.2924}           & \multicolumn{1}{c|}{{0.3438}}             & {0.6174}            & {0.6655}             & {0.3903}          & {0.4033}            \\
\multicolumn{1}{l|}{{\textbf{SGL+Attr-A}}}   & {0.4356}          & {0.535}          & {0.2489}         & \multicolumn{1}{c|}{{0.2779}}          & {0.5501}        & {0.5924}        & {0.3528}        & {0.3646}         \\
\multicolumn{1}{l|}{{\textbf{SGL+Attr-B}}}   & {0.503}          & {0.6466}          & {0.2782}         & \multicolumn{1}{c|}{{0.3214}}          & {0.6079}        & {0.6474}        & {0.3946}        & {0.4053}         \\
\multicolumn{1}{l|}{\textbf{SGL}}      & \underline{0.5545}   & \underline{0.7211}  & 0.3128        & \multicolumn{1}{c|}{0.3633}         & \underline{0.6355} & \textbf{0.6818} & \textbf{0.4166} & \textbf{0.4291} \\
\multicolumn{1}{l|}{\textbf{SSL-DNN}} & 0.5444          & 0.7116         & \underline{0.3139}   & \multicolumn{1}{c|}{\underline{0.3647}}    & 0.6236        & 0.6705        & 0.3999        & 0.4124        \\
\multicolumn{1}{l|}{\textbf{IFL}}     & \textbf{0.5549} & \textbf{0.7230} & \textbf{0.3190} & \multicolumn{1}{c|}{\textbf{0.3699}} & \textbf{0.6400} & \underline{0.6814}        & \underline{0.4088}        & \underline{0.4200}          \\ \hline
\end{tabular}
}
}
\end{center}
\end{table*}
\vspace{5pt}
\noindent\textbf{Baselines.} We compare the proposed IFL framework with the following five representative recommender models:
\begin{itemize}[leftmargin=*]
    \item \textbf{FM} \cite{rendle2010factorization}, \textbf{NFM} \cite{he2017nfm}, and \textbf{DeepFM} \cite{guo2017deepfm} are powerful feature-based recommenders, which blindly utilize all features by feature interaction modeling to generate recommendations.
    
     \item \textbf{SSL-DNN}~\cite{yao2021self} is a multi-task-based self-supervised learning recommendation framework with two towers of DNN. It adopts 
     a feature dropout strategy for data augmentation. {Here, we utilize correlated feature masking (CFM) for feature dropout due to its better performance compared to random feature masking}~\cite{yao2021self}, {and SSL-DNN shares the same base neural network and learnable parameters as IFL}.
     
    
    \item \textbf{SGL}~\cite{wu2021self} is a self-supervised learning method for graph-based collaborative filtering~\cite{Liu2021IMP_GCN}, which only utilizes the feature of the user ID and item ID. It conducts data augmentation by node dropout and edge dropout.
    
    \item \textbf{SGL+Attr-A} and \textbf{SGL+Attr-B} are two variants of SGL that utilize side information. We add the features for SGL+Attr-A and SGL+Attr-B before and after the graph propagation, respectively. 

    \item {\textbf{AutoDeepFM}~\cite{autofis} is a method that can automatically discover and remove redundant feature interactions based on informative feature selection. It identifies the informative cross features by assigning learnable weights to feature interactions.
    }
    
    \item {\textbf{AutoFIS+CFM} adopts the informative feature selection into SSL-DNN. Learnable weights for features are used by SSL-DNN to discover the redundant features, and the model will be re-trained with only informative features.
    }
\end{itemize}

\vspace{5pt}
\noindent\textbf{Evaluation.} 
Following the common setting of existing work, we adopt the full-ranking protocol to evaluate the top-K recommendation performance with the averaged Recall@K and NDCG@K as the metrics. We set $K=\{50, 100\}$ for Meitaun and $K=\{10,20\}$ for XING.  We compute the metrics on both the IID and OOD testing sets, respectively.


\vspace{5pt}
\noindent\textbf{Hyper-parameter Settings.} 
We fix the embedding size $64$ for all methods. Then, we tune the hyper-parameters on the validation sets with the following strategies:

\begin{itemize}[leftmargin=*]
   \item 
   For \textbf{FM}, \textbf{NFM}, \textbf{DeepFM}, and \textbf{AutoDeepFM}, we tune their learning rate, DNN (implemented as MLP) hidden sizes, dropout ratio, L2-regularization coefficient, batch normalization, and negative sampling number in the ranges of $\{0.01,0.001\}$, $\{[32],[64]\}$, $\{0.4,0.5,0.6\}$, $\{1e\text{-}4,1e\text{-}3,0.1,0.2\}$, $\{0,1\}$, $\{1,5,10,20\}$, respectively.
    
    \item For \textbf{SSL-DNN} and \textbf{AutoFIS+CFM}, we search the learning rate, DNN (MLP) hidden sizes, dropout ratio, strength coefficient of SSL loss, and temperature coefficient of SSL loss for smoothing in the ranges of $\{0.01,0.005\}$, $\{[none],[64]\}$, $\{0.1,0.2\}$, $\{0.1,0.5,1\}$, and $ \{0.1,0.3,0.5,0.7,1\}$, respectively. Here, $[none]$ represents that the DNN only contains a fully-connected layer.
    
    \item For \textbf{SGL}, \textbf{SGL+Attr-A}, and \textbf{SGL+Attr-B}, we adopt the version named SGL-ED and follow its default setting to set the number of graph layers to three and the learning rate to $0.001$. Then, we tune the L2-regularization coefficient, strength coefficient of SSL loss, and temperature of NCE loss for smoothing in the ranges of $\{0.0001,0.1\}$, $\{0.1,0.2,0.5\}$ and $\{0.1,0.3,0.5,0.7,1.0\}$, respectively.
    
    
    
    \item For the proposed \textbf{IFL}, we tune the hyper-parameters $\alpha$, $\beta$, $\tau$ in the loss, the number of environments, and the learning rate within the ranges of $\{0,0.1,0.3,0.5,0.7,1.0\}$, $\{0,0.0001,0.001,0.01,0.1,1.0\}$, $\{0.1,0.3,0.5,\\0.7,1.0\}$, $\{1,2,4\}$ and $\{0.01,0.05\}$, respectively.
    
\end{itemize}

\subsection{Overall Performance (RQ1)}
We compare the recommendation performance in both IID and OOD settings between IFL and the baselines. The results are shown in Table~2, where we have the following observations:
\begin{itemize}[leftmargin=*]
    
    \item In the OOD testing setting, the proposed IFL outperforms all baselines, including SSL-based methods, on both two datasets. To achieve a good OOD generalization, a model should eliminate the impact of spurious correlation in the training set. Thus, this result can verify the effectiveness of our proposal --- spurious feature mask learning and mask-guided data augmentation --- in identifying the spurious features and blocking the bad effect of the corresponding spurious correlations for SSL. 
    
    \item In the IID testing setting, our IFL at least achieves comparable performance to SOTA baselines. Together with its good OOD performance, this result shows that IFL can achieve good OOD generalization ability without sacrificing IID prediction.

    \item When moving from the IID testing setting to the OOD testing setting, all methods show sharp performance drops. This is a normal phenomenon, since there are huge distribution shifts from the training to the OOD testing set. Although our IFL is expected to achieve OOD performances by removing the spurious correlation in training, we cannot expect it to have no performance drops since 1) the OOD set is not an unbiased set and 2) the data distribution itself also influences the model learning and further affects model performances. 
    
    \item Among all baselines, SGL usually achieves better performance. This can be attributed to its good ability to model high-order interactions via graph learning and its effective graph-based data augmentation. Meanwhile, it is less possible to be affected by the spurious features with only ID information, achieving better OOD performances. However, SGL-Attr-A and SGL-Attr-B perform worse than SGL, which may be attributed to the noise introduced by spurious features. 
    Therefore, it verifies that the superior performance of our proposed IFL comes from the invariant features learned by the feature mask mechanism. 
    Moreover, the better performance of SSL-DNN compared to AutoFIS+CFM and the relatively poor performance of SSL-DNN compared to IFL and SGL indicates the necessity to deal with the spurious correlation issues for SSL and the effectiveness of IFL again.

    
    
\end{itemize}

\subsection{Ablation Study (RQ2)}
Feature mask learning is key to identifying the spurious features and further helping to remove these features. In this section, to verify the importance of its key components, we conduct the following ablation studies: 



\begin{itemize}[leftmargin=*]

    \item The variances loss in the feature mask learning is the key to pushing the feature mask mechanism to identify the invariant features and discard the spurious features. Therefore, we next study its importance. Specially, we compare the IFL with the variance loss and a variant of IFL without the loss. Fig.~\ref{fig:variance} shows the results. We can find: 1) in the IID testing setting, the two models show similar performance; 2) but in the OOD testing setting, the performance of the IFL without the variance loss decreases obviously, compared to the IFL with the loss. This shows that the variance loss is necessary for helping the feature masking mechanism identify and discard the spurious features.
    
    
    \begin{figure}[t]
    \vspace{-0.5cm}
    \setlength{\abovecaptionskip}{0cm}
    \setlength{\belowcaptionskip}{0cm}
      \centering 
      \hspace{-1.4in}
      \subfigure{
        \includegraphics[width=1.6in]{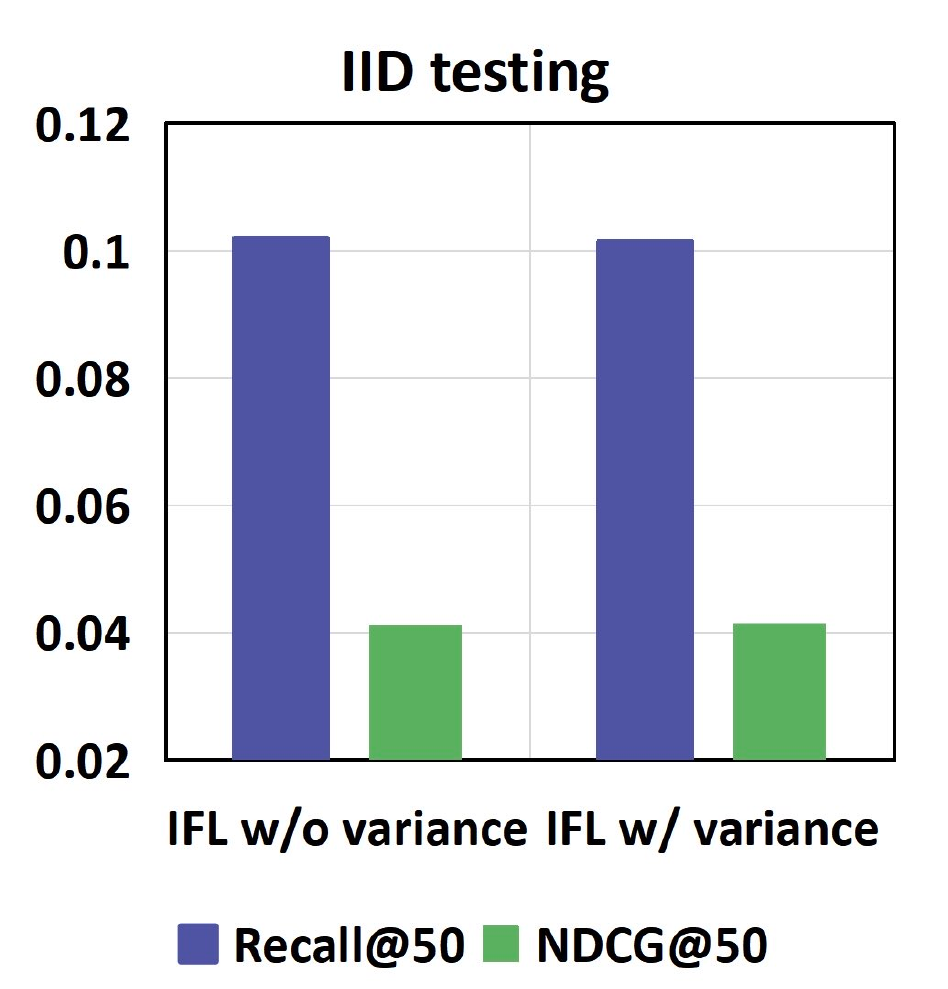}} 
      \hspace{-0.2in}
      \subfigure{
        \includegraphics[width=1.6in]{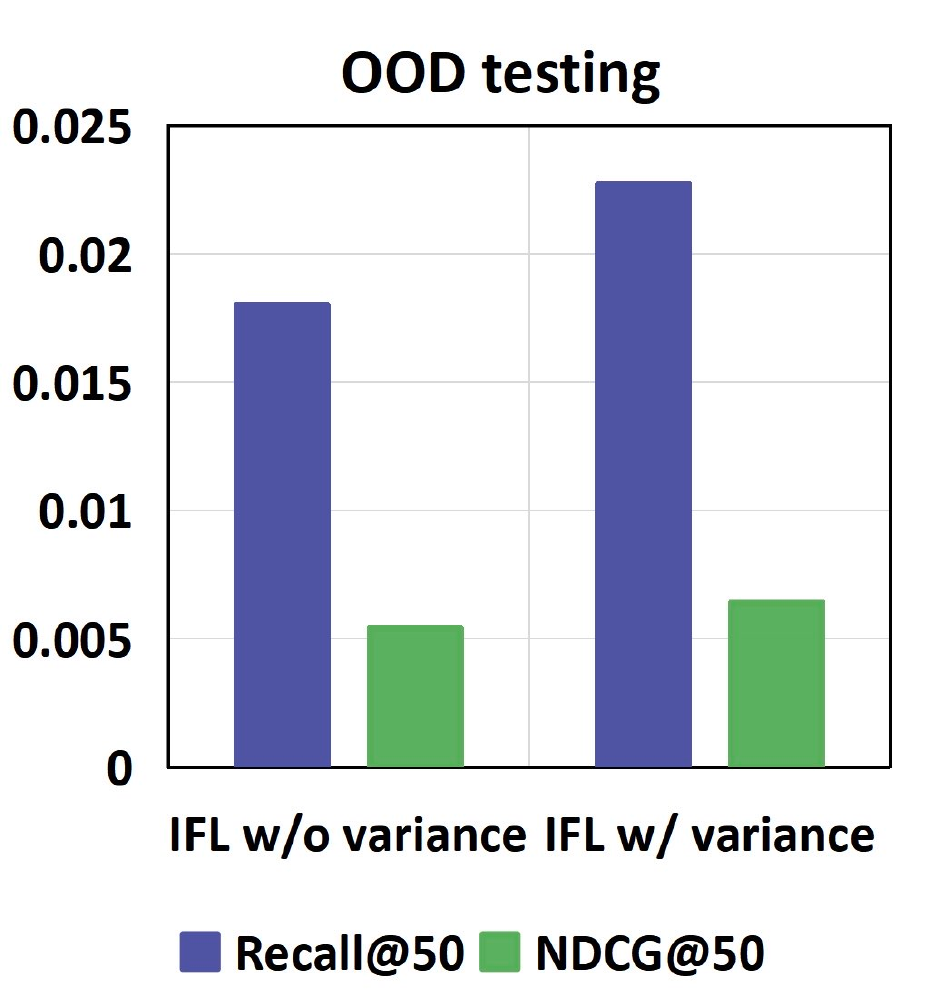}} 
      \hspace{-1.5in} 
      \vspace{-0.1cm}
      \caption{The performance of IFL with (w/, $\beta=0.01$) or without (w/o, $\beta=0$) variance loss on IID and OOD over Meituan.}
      \label{fig:variance}
    \end{figure}

    \item To verify the importance of the proposed spurious feature mask mechanism,  we compare the IFL with (w/) and the IFL without (w/o) the feature mask mechanism. The results are shown in Fig.~\ref{fig:mask}. From the figure, we find that the IFL without the feature mask shows poorer performances on both the IID testing and the OOD testing. The results of the OOD testing indicate that it is necessary to remove the effects of spurious features for OOD generalization, and the proposed feature mask mechanism effectively discards the effects.
    Meanwhile, the results for the IID testing show that removing spurious features could benefit the IID prediction. The possible reason is that: the spurious features could be non-necessary features for the IID prediction. Such non-necessary features could introduce noise to the model learning as discussed in~\cite{autofis}. Note that directly removing such features is not helpful for OOD generalization, which can be proved by the fact that IFL without variance loss also conducts the feature mask mechanism, but shows poor performances (\cf Fig.~\ref{fig:variance}).  
    
    
    
    \begin{figure}[t]
    \vspace{-0.5cm}
    \setlength{\abovecaptionskip}{0cm}
    \setlength{\belowcaptionskip}{0cm}
      \centering 
      \hspace{-1.4in}
      \subfigure{
        \includegraphics[width=1.6in]{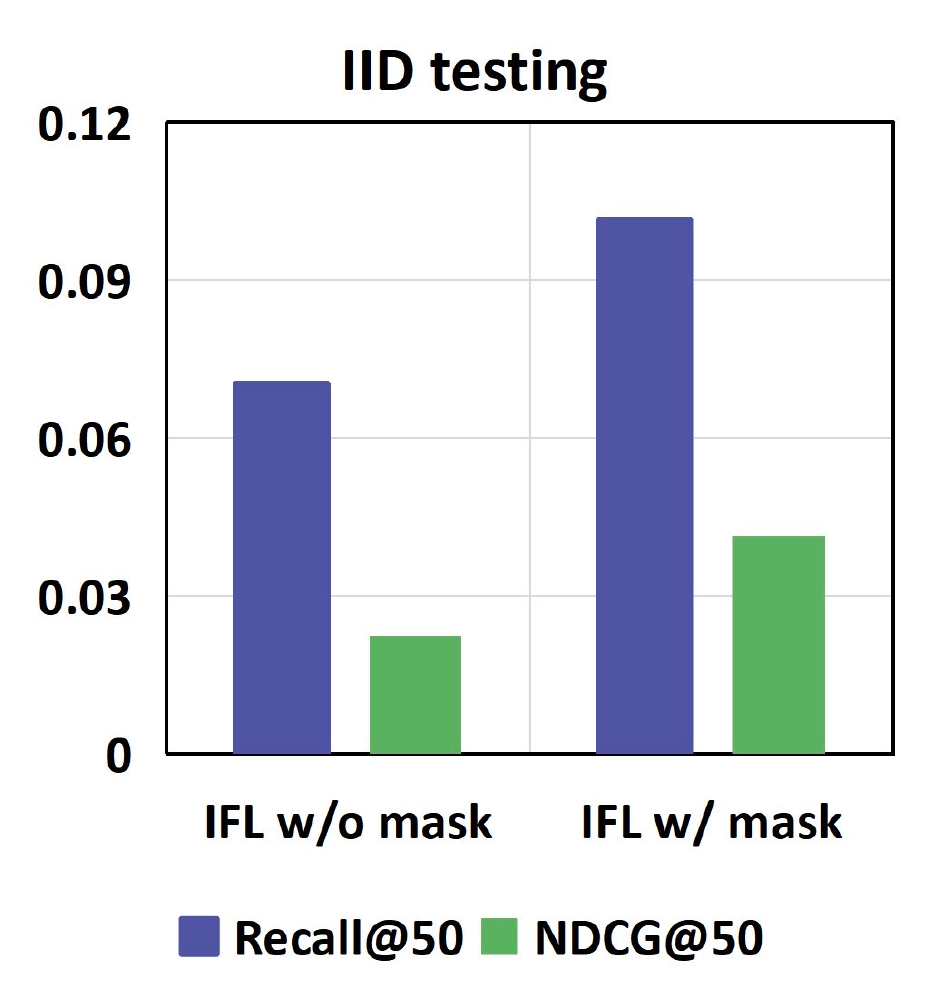}} 
      \hspace{-0.2in}
      \subfigure{
        \includegraphics[width=1.6in]{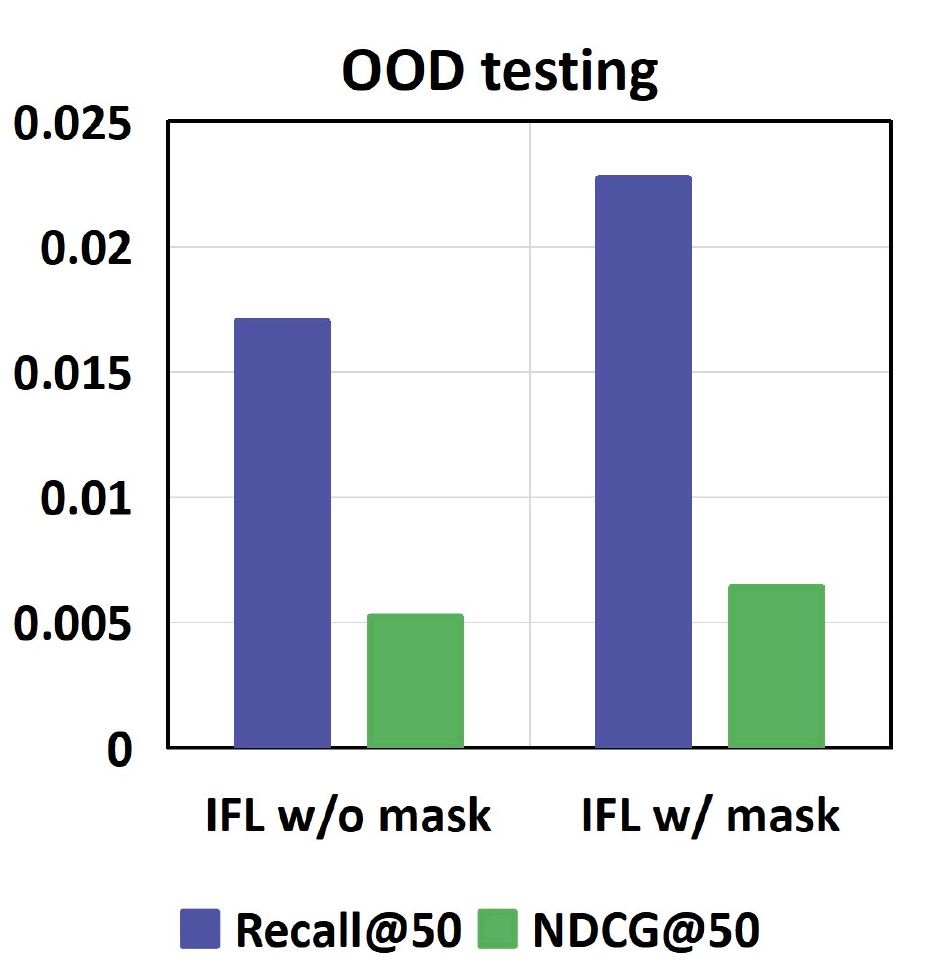}} 
      \hspace{-1.5in} 
      \vspace{-0.1cm}
      \caption{The performance of IFL with (w/) or without (w/o) performing the feature mask mechanism under the IID and OOD testing settings on Meituan.}
      \label{fig:mask}
    \end{figure}

\end{itemize}


\begin{figure*}[ht]
\vspace{-0.5cm}
\setlength{\abovecaptionskip}{-0.20cm}
\setlength{\belowcaptionskip}{0cm}
  \centering 
  \hspace{0in}
  \subfigure{
    \includegraphics[width=1.9in]{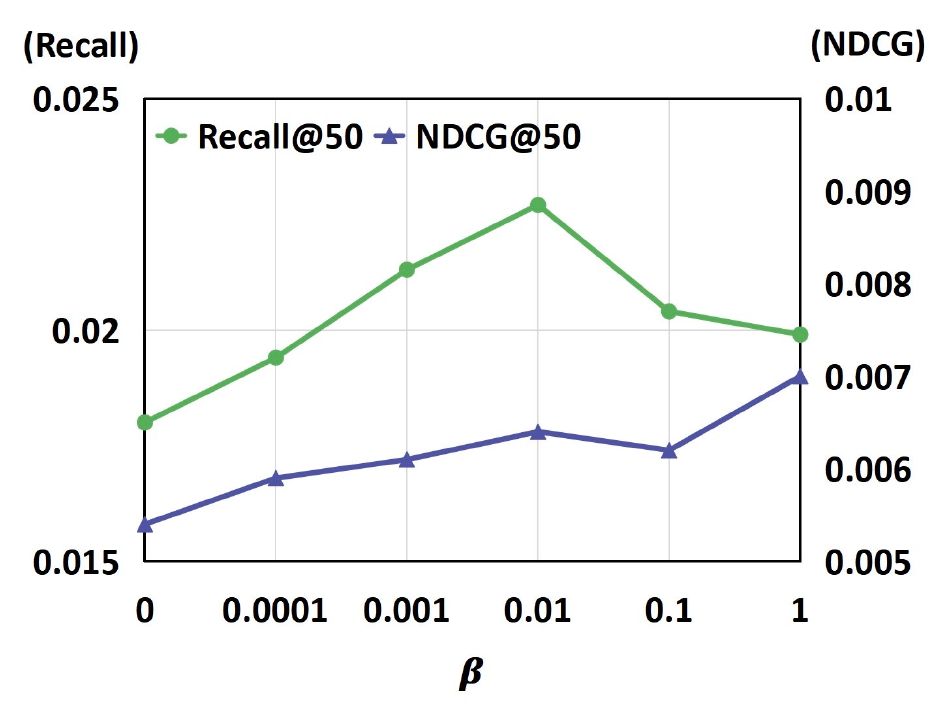}} 
  \hspace{0.1in}
  \subfigure{
    \includegraphics[width=1.9in]{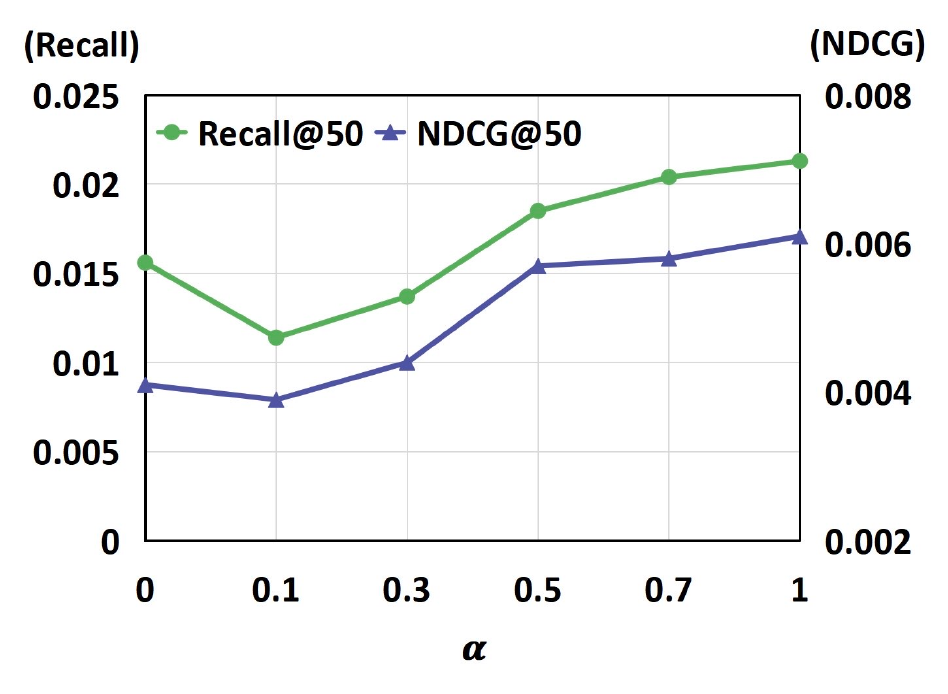}} 
  \hspace{0.1in}
  \subfigure{
    \includegraphics[width=1.9in]{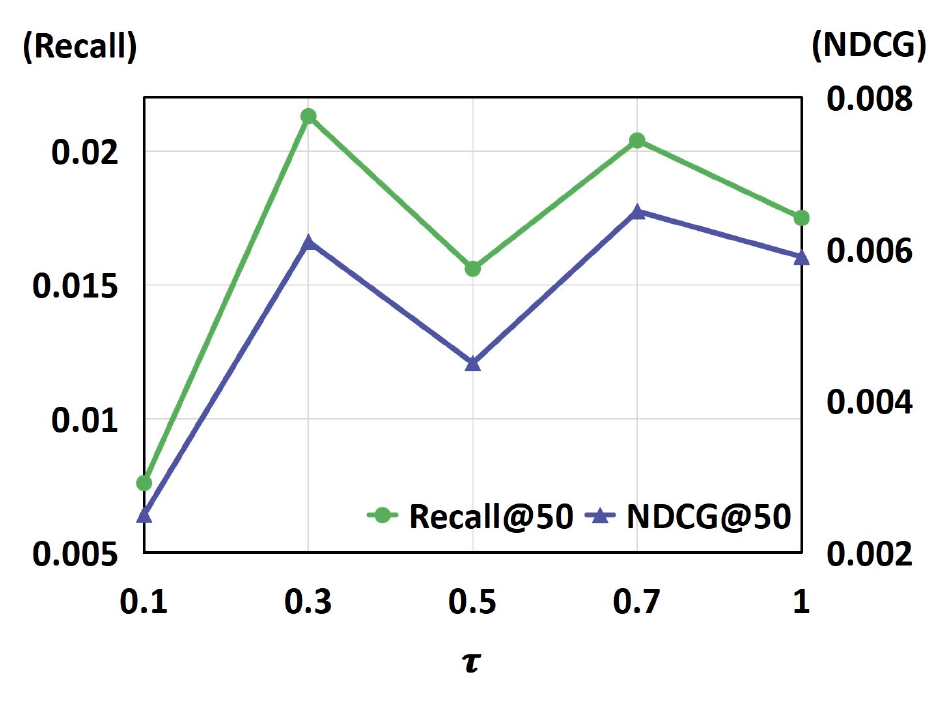}} 
  \hspace{0in} 
  \vspace{-0.1cm}
\caption{Hyper-parameter influences on the OOD testing performances for Meituan, regarding the hyper-parameters $\beta$ to control the influence of variance loss, $\alpha$ to adjust the weight of SSL loss and $\tau$ in SSL loss for smoothing.}
  \label{fig:sensityvity-ood}
  \vspace{-0.50cm}
\end{figure*}

\begin{figure*}[ht]
\setlength{\abovecaptionskip}{-0.20cm}
\setlength{\belowcaptionskip}{0cm}
  \centering 
  \hspace{0in}
  \subfigure{
    \includegraphics[width=1.9in]{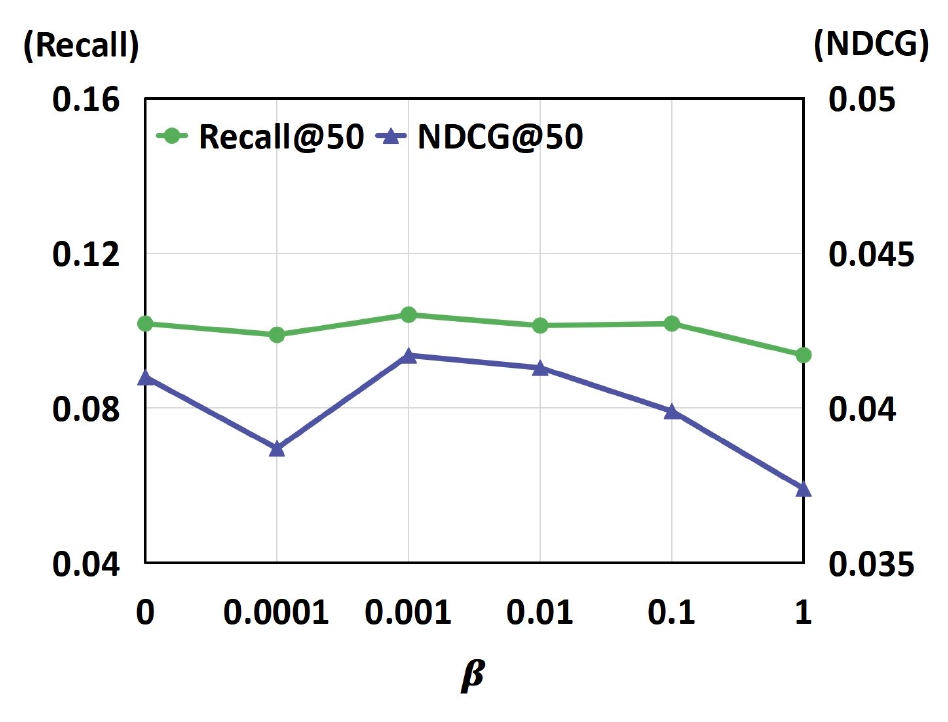}} 
  \hspace{0.1in}
  \subfigure{
    \includegraphics[width=1.9in]{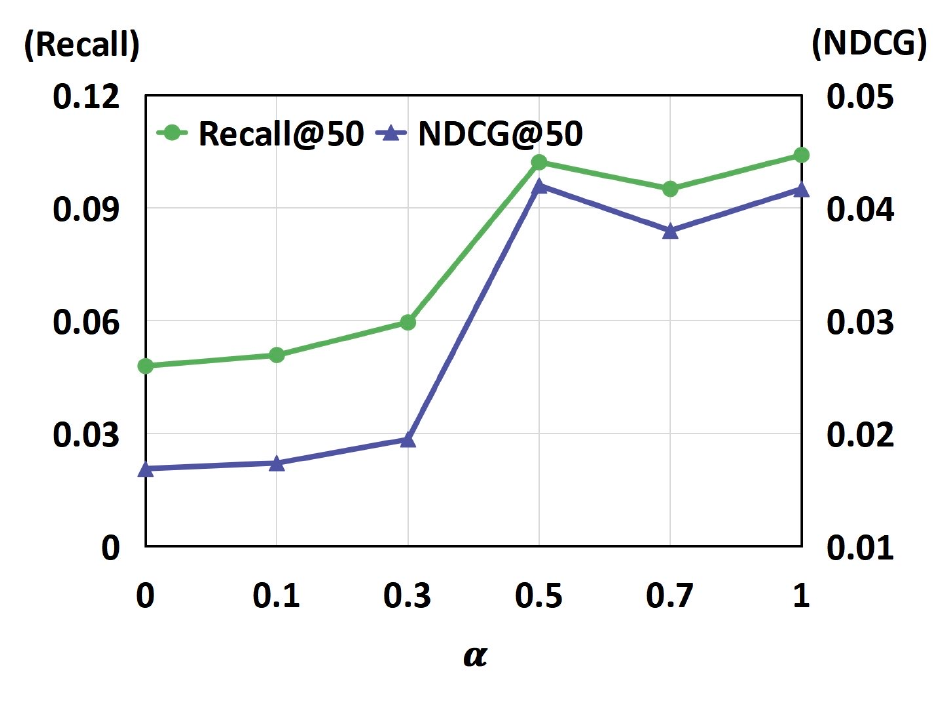}} 
  \hspace{0.1in}
  \subfigure{
    \includegraphics[width=1.9in]{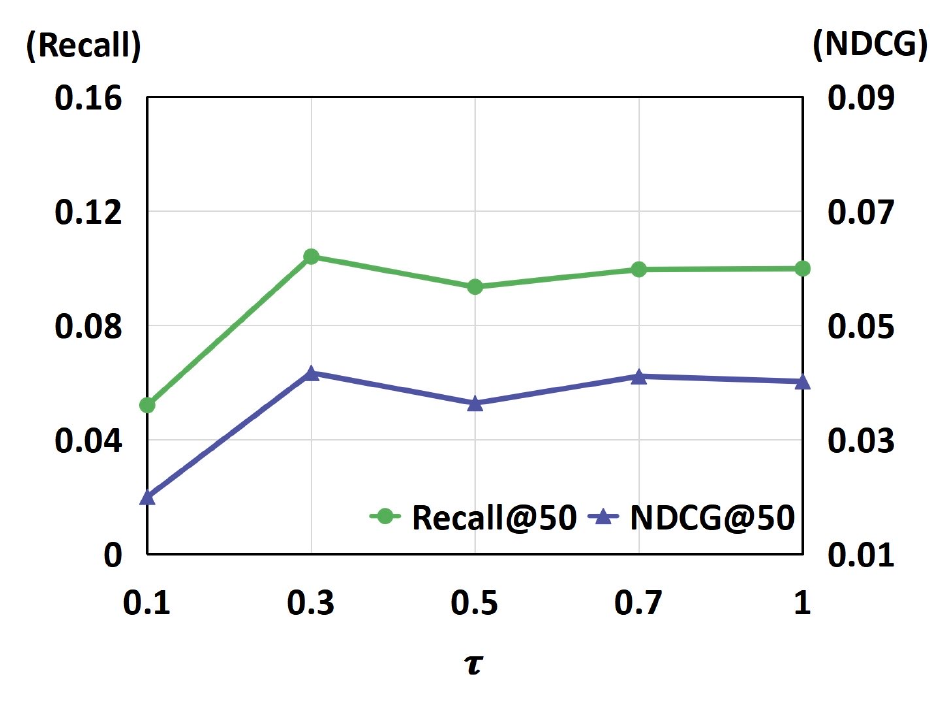}} 
  \hspace{0in} 
  \caption{
   The influences of hyper-parameters on the IID testing performances for Meituan. There are three hyper-parameters --- the hyper-parameters $\beta$ to control the influence of variance loss, $\alpha$ to adjust the weight of SSL loss and $\tau$ in SSL loss for smoothing.
  }
  \label{fig:sensityvity-iid}
  \vspace{0cm}
\end{figure*}

\subsection{Hyper-parameter Analysis (RQ3)}
To analyze the influences of different hyper-parameters of IFL on the recommendation performance, we conduct a series of experiments on Meituan by varying these hyper-parameters. Note that when studying a hyper-parameter, we fix other hyper-parameters as the optimal in Table~2. 
For the hyper-parameters $\beta$ to control the influence of variance loss, $\alpha$ to adjust the weight of SSL loss, and $\tau$ in SSL loss for smoothing, we vary them in the ranges of $\{0,1e-4,1e-3,1e-2,0.1,1\}$, $\{0,0.1,0.3,0.5,0.7\}$ and $\{0.1,0.3,0.5,0.7\}$, respectively. Fig.~\ref{fig:sensityvity-ood} and Fig.~\ref{fig:sensityvity-iid} show the results for them on OOD and IID testing settings, respectively. For the hyper-parameter $C$, we vary it in the range of $\{1,2,4\}$, and Table~3 shows the corresponding results. From the figures and the table, we have the following observations:



\begin{itemize}[leftmargin=*]

\item For $\beta$, according to the sub-figures on the left of Fig.~\ref{fig:sensityvity-ood} and Fig.~\ref{fig:sensityvity-iid}, we find that $\beta$ has a slight influence on the IID testing performance. While in the OOD testing setting, the performance roughly shows a decreasing trend when $\beta$ decreases. When $\beta$ decreases to $0$, \ie no variance loss to push the feature mask mechanism to discard the spurious features, our IFL cannot outperform the normal SSL method SGL (\cf Table~2). Thus, when tuning the $\beta$ for IFL, we do not need to pay much attention to the IID performance, and we need to set a relatively large $\beta$ to achieve better OOD performance.


\item For $\alpha$, as the middle sub-figures of Fig.~\ref{fig:sensityvity-ood} and Fig.~\ref{fig:sensityvity-iid} show, the OOD and IID testing performances show similar increasing trends when $\alpha$ increases. Thus, when tuning it, we do not need to consider the trade-off between the OOD and IID performance. Besides, we note that when the $\alpha=0$, \ie the SSL part is disabled for IFL, the OOD testing performances of IFL decrease sharply, but IFL can still outperform other non-SSL baselines in the OOD testing setting (\cf Table~2). This shows that SSL is important for IFL but the proposed feature mask learning of IFL is also useful for non-SSL models.

\item Regarding $\tau$, according the sub-figures on the right of Fig.~\ref{fig:sensityvity-ood} and Fig.~\ref{fig:sensityvity-iid}, we find that $\tau$ has sensitive influences on both the IID and OOD testing performance, which is similar to previous work~\cite{wu2021self}.
Fortunately, the OOD and IID testing performances show very similar fluctuating trends, making hyper-parameter tuning relatively easy. 


\item Regarding $C$, as Table~3 shows, with $C$ changing, the IID performance remains relatively stable while the OOD performance is hugely affected by $C$. Since the quality and the number of the environments usually have huge influences on the effectiveness of invariant learning~\cite{HRM}, we need to tune it carefully.

\end{itemize}

\begin{table}[t]
\setlength{\abovecaptionskip}{0cm}
\setlength{\belowcaptionskip}{-1cm}
\setlength{\tabcolsep}{1.5mm}{
\caption{The influences of the hyper-parameter $C$ to adjust the number of environments on Meituan.}
}
\begin{center}
\setlength{\tabcolsep}{1.5mm}{
\resizebox{0.48\textwidth}{!}{
\begin{tabular}{c|cc|cc}
\hline
\multicolumn{1}{l|}{} & \multicolumn{2}{c|}{\textbf{OOD testing}} & \multicolumn{2}{c}{\textbf{IID testing}} \\
\multicolumn{1}{l|}{$C$} & \textbf{Recall@50}   & \textbf{NDCG@50}   & \textbf{Recall@50}   & \textbf{NDCG@50}  \\ \hline
\textbf{1}            & 0.0180                & 0.0054             & 0.1018               & 0.0410             \\
\textbf{2}            & \textbf{0.0213}      & \textbf{0.0061}    & \textbf{0.1041}      & \textbf{0.0417}   \\
\textbf{4}            & 0.0189               & 0.0056             & 0.1018               & 0.0410             \\ \hline
\end{tabular}
}
}
\end{center}
\label{table:sensitivity_env}
\end{table}

\subsection{Case Study}
To illustrate how IFL blocks the negative effect of spurious features, we analyze the feature masks and examine the interaction distributions. 
From Fig.~10
, we find that some user features are recognized as spurious features with the feature mask value as zero. Thereafter, we investigate the interaction distributions on XING to verify the effectiveness of the feature mask.

\begin{figure}[tb]
    \setlength{\abovecaptionskip}{-0.2cm}
    \setlength{\belowcaptionskip}{-0.2cm}
    \centering
    \includegraphics[scale=0.45]{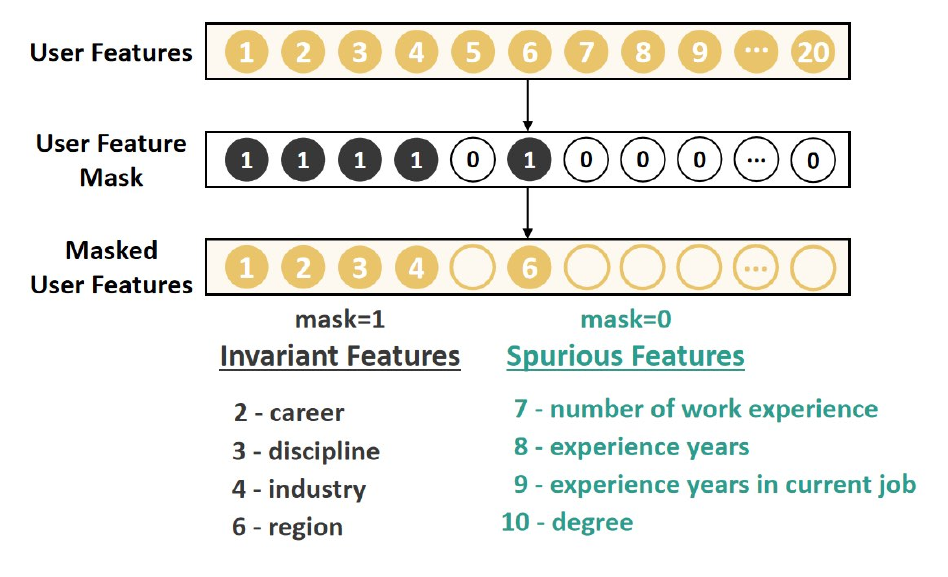}
    {\caption{User feature mask learned by IFL in XING. The mask value of 1 indicates the invariant feature, while that of 0 indicates the spurious feature.}}
    \label{fig:case1}
\end{figure}

Here, we demonstrate two examples of the discovered spurious user features (i.e., experience years and experience years in the current job). According to the figure on the left in Fig.~11
, among all full-time jobs, users with 4 and 6 years of work experience are more likely to have positive interactions, while those with 5 and 7 years are more likely to have negative interactions. However, based on expert knowledge, the preference of users with only one year gap in years of work experience over full-time jobs should not be significantly different, which makes users’ number of years of experience a spurious feature. Similarly, from the graph to the right of Fig.~11
, user preference for jobs in Austria is drastically influenced by the work experience in the current job. In fact, this is irrational; thus, the years of experience of users in their current job is also a spurious feature. In prediction, these identified spurious features are well shielded by the feature mask, which intuitively explains the superior generalization ability of IFL.


\begin{figure}[t]
    \vspace{-0.5cm}
    \setlength{\abovecaptionskip}{0cm}
    \setlength{\belowcaptionskip}{0cm}
      \centering 
      \hspace{-1.4in}
      \subfigure{
        \includegraphics[width=1.35in]{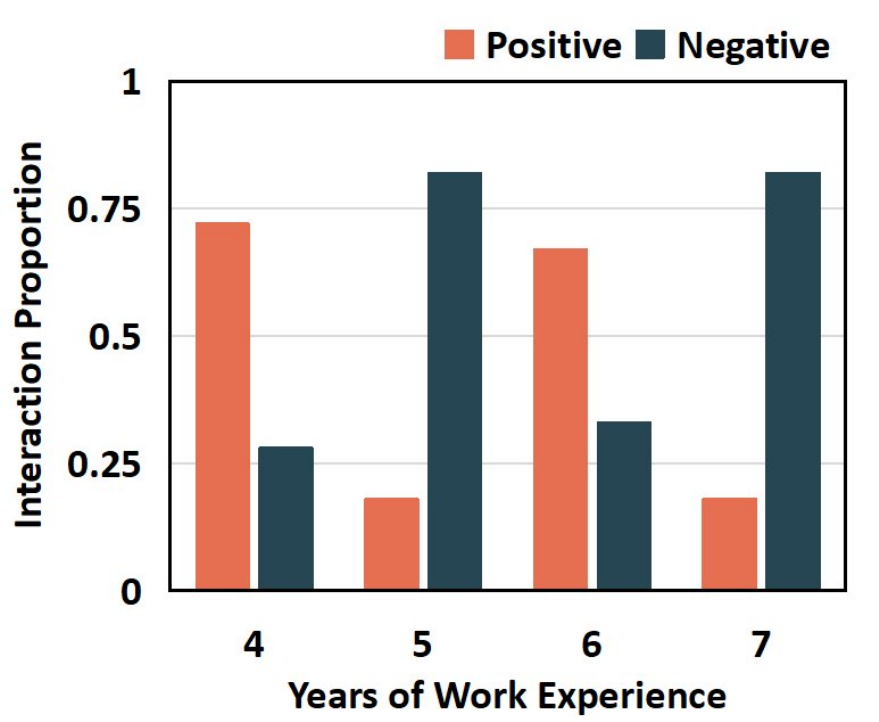}} 
      \hspace{-0.1in}
      \subfigure{
        \includegraphics[width=1.6in]{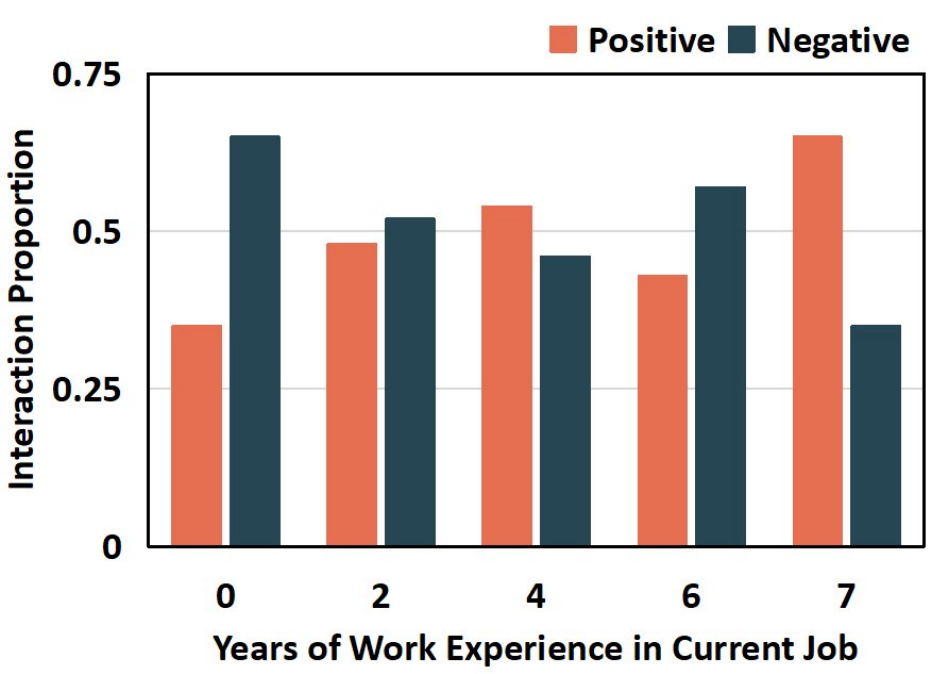}} 
      \hspace{-1.5in} 
      \vspace{-0.1cm}
      {\caption{The figure to the left presents the distribution of positive/negative interactions over users' years of work experience among all full-time jobs. The figure to the right demonstrates the distribution of positive/negative interactions over users' years of work experience in their current job among all jobs in Austria.}}
      \label{fig:case}
\end{figure}
\section{Related Work}\label{sec:related_work}

\subsection{Neural Recommendation}
With the rise of deep learning in machine learning, neural recommendation is becoming more and more flourishing~\cite{Wule-NeuralRec,alqwadri2021app,Liu2022SDG} and usually can surpass traditional recommenders~\cite{Wule-NeuralRec}. Technically, we can categorize the existing work into two lines. The first line is to develop neural recommenders based on various deep neural networks~\cite{Wule-NeuralRec}, including work based on MLP~\cite{MLP-youneed,ncf}, convolutional neural networks~\cite{caser}, self-attention~\cite{bert4rec}, \textit{etc}. 
Another line of efforts models recommendation data with different graphs, \eg bipartite graph~\cite{lightgcn}, knowledge graph~\cite{KGIN}, and hypergraph~\cite{hyperG-rec}, and then designs recommenders with graph neural networks.
However, these methods are trained in a supervised-learning manner, and the supervised signals (\ie interactions data) are extremely sparse compared to the overall sample space, limiting the effectiveness of neural recommenders. Differently, we utilize self-supervised learning, which has potential to overcome this drawback.  

\subsection{Self-supervised Learning}
Regarding SSL, contrastive models are the most related, which learn to compare samples through a Noise Contrastive Estimation (NCE) objective~\cite{wu2021self}.
Some work focuses on modeling the contrast between the local part of a sample and its global context~\cite{SSL-LG-1}, and other work performs comparisons between different views of samples ~\cite{she2021contrastive,SSL-G2G}.
Benefiting from its relatively lower dependency on labeled data, SSL also receives huge attention in recommendation~\cite{SSL-rec-survey}.
Many attempts have been proposed for various neural recommendations to overcome the data sparsity challenge~\cite{wu2021self,SSL-rec-survey,yao2021self}. 
For example, SGL~\cite{wu2021self} applies the SSL to the graph recommender model based on node and edge dropout.
Although SSL has become popular in many areas and has become the SOTA for personalized recommendation~\cite{wu2021self,yao2021self}, 
SSL might capture spurious correlations, since it blindly discovers correlation relationships with the self-discrimination task, resulting in poor generalization ability. In this work, we try to overcome this drawback by combining invariant learning. 

\subsection{Causal Recommendation}
Data-driven recommender systems achieve great success in large-scale recommendation scenarios~\cite{Wule-NeuralRec}. However, recent work finds that they face various biases~\cite{PDA,IPW-saito,wang2021deconfounded}, unfairness~\cite{fairness}, and low OOD generalization ability issues~\cite{wenjie-ood}. The reason is thought of as the lack of modeling causality to avoid capturing spurious correlations~\cite{PDA,wenjie-ood}. Many efforts try to incorporate causality into neural recommendations to overcome these drawbacks~\cite{PDA,wenjie-ood,IPW-saito}. There are mainly two types of work. The first line of research is based on the potential outcome framework~\cite{IPW-saito, rubin2005causal}, where IPS~\cite{IPW-saito} and doubly robust~\cite{DR} are utilized to achieve unbiased recommendation. Another line of research is based on the structural causal model~\cite{pearl2009causality,PDA,wenjie-clickbait,wenjie-ood}.
The existing efforts usually analyze the causal relationships with causal graphs and then estimate the target causal effect with the intervention~\cite{PDA} 
or counterfactual inference~\cite{wenjie-ood,wenjie-clickbait} for debiasing, fairness, or OOD generalization. Nevertheless, all previous methods do not consider dealing with the spurious correlation issues for SSL. Besides, to the best of our knowledge, existing work in recommendation does not take invariant learning~\cite{IRM,HRM} to remove spurious correlations.




\section{Conclusion}\label{sec:conclusion}
In this work, we inspected spurious correlations in the SSL recommendation. To improve the generalization ability of SSL recommender models, we proposed the IFL framework to exclude spurious features and leverage the invariant features for recommendation. Specifically, we considered a feature mask mechanism to automatically recognize spurious features and utilized mask-guided contrastive learning to block the harmful effect transmission from spurious features to invariant features. Extensive experiments validate the superiority of IFL in mitigating spurious correlations and enhancing the generalization ability of SSL models.  

This work makes the initial attempts to mitigate the effect of spurious correlations on the SSL recommendation. In future work, one promising direction is learning the personalized feature mask mechanism to discover user-specific invariant features. Additionally, to achieve better generalization performance, it is worth improving the approaches to automatically divide environments with distribution shifts. 






\bibliography{mir-article}

\begin{figure}[h]%
\centering
\includegraphics[width=0.1\textwidth]{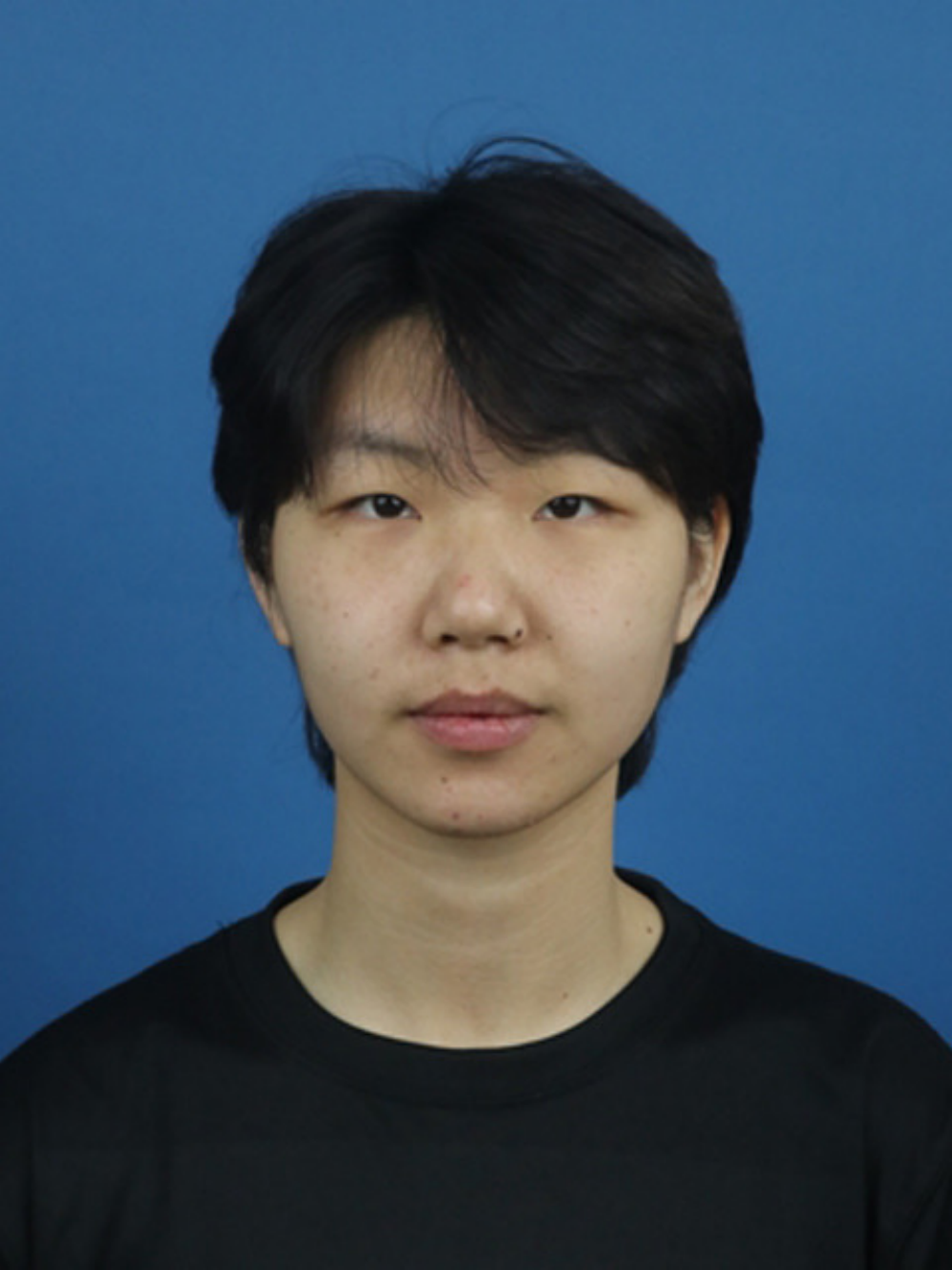}
\vspace{-0.5cm}
\end{figure}
\noindent{\bf Xinyu Lin} is a Master student in the School of Science, National University of Singapore. She received the B.E. degree from the school of Control Science and Engineering, Shandong University in 2021. Her research interests cover causal recommendation, causal representation learning, and multimedia analysis. \\
E-mail: xylin1028@gmail.com \\
ORCID: 0000-0002-6931-3182
\vspace{10pt}

\begin{figure}[h]%
\centering
\includegraphics[width=0.1\textwidth]{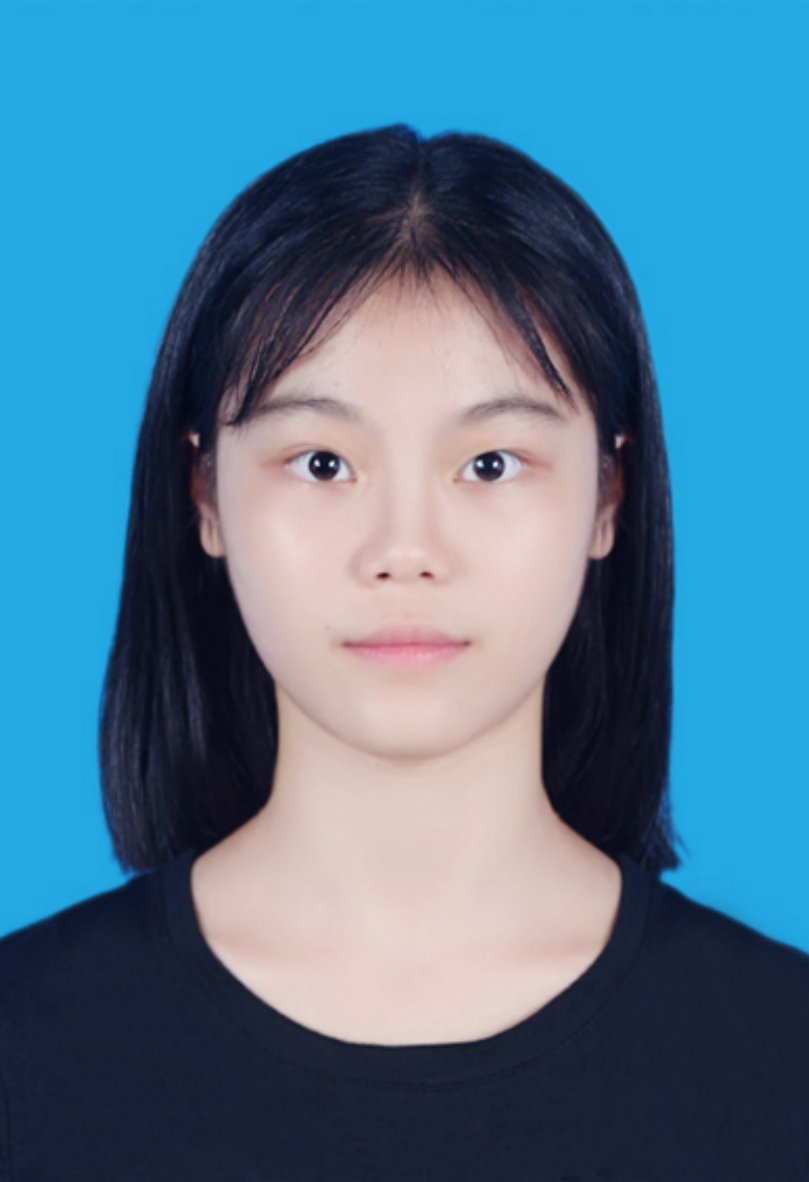}
\vspace{-0.5cm}
\end{figure}
\noindent{\bf Yiyan Xu} is a Master student in the School of Data Science, University of Science and Technology of China (USTC), supervised by Prof. Fuli Feng. She received the B.S degree from the QianWeiChang College, Shanghai University in 2022. Her research interest lies in the recommender system, graph neural networks and causal inference. \\
E-mail: yiyanxu24@gmail.com \\
ORCID: 0000-0002-5937-7289
\vspace{10pt}

\begin{figure}[h]%
\centering
\includegraphics[width=0.1\textwidth]{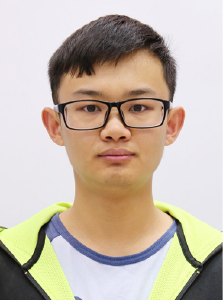}
\vspace{-0.5cm}
\end{figure}
\noindent{\bf Wenjie Wang} is a Ph.D. student in the School of Computing, National University of Singapore. He received the B.E. degree from the School of Computer Science and Technology, Shandong University in 2019. His research interests cover causal recommendation, data mining, and multimedia. His publications have appeared in top conferences and journals such as SIGIR, KDD, WWW, and TIP. Moreover, he has served as the PC member and reviewer for the top conferences and journals including TKDE, TOIS, SIGIR, AAAI, ACMMM, and WSDM.\\
E-mail: wenjiewang96@gmail.com (Corresponding author) \\
ORCID: 0000-0002-5199-1428
\vspace{10pt}

\begin{figure}[h]%
\centering
\includegraphics[width=0.1\textwidth]{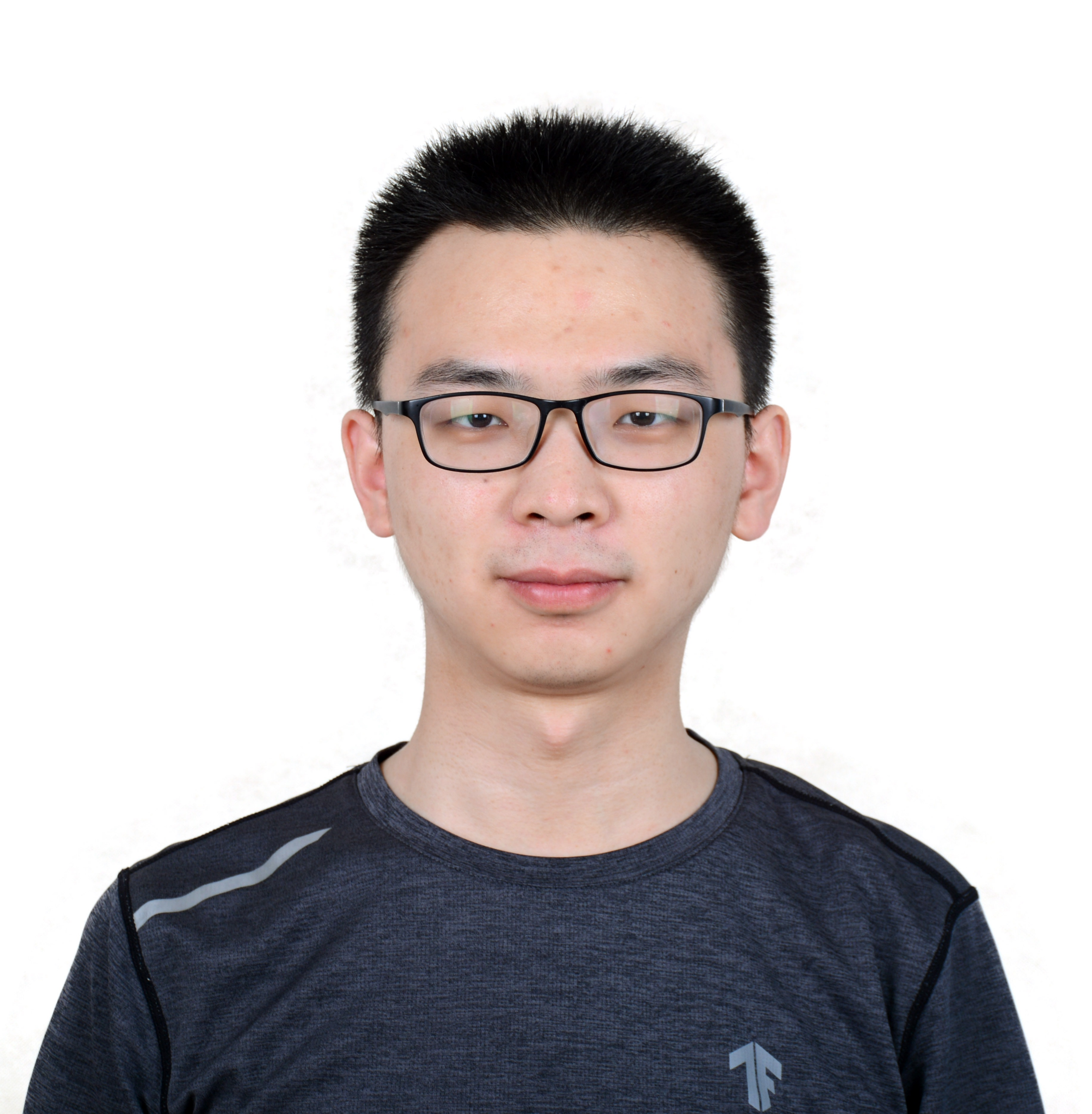}
\end{figure}
\noindent{\bf Yang Zhang} is a Ph.D. student in the School of Information Science and Technology, University of Science and Technology of China (USTC), supervised by Prof. Xiangnan He. He received his B.E. degree from the USTC. His research interest lies in the recommender system and causal inference. He has two publications in the top conference SIGIR. His work on the causal recommendation has received the Best Paper Honorable Mention in SIGIR 2021. He has served as the PC member and reviewer for the top conferences and journals including TOIS, TIST, ICML-PKDD, AAAI and WSDM.\\
E-mail: zy2015@mail.ustc.edu.cn


\begin{figure}[h]%
\centering
\includegraphics[width=0.1\textwidth]{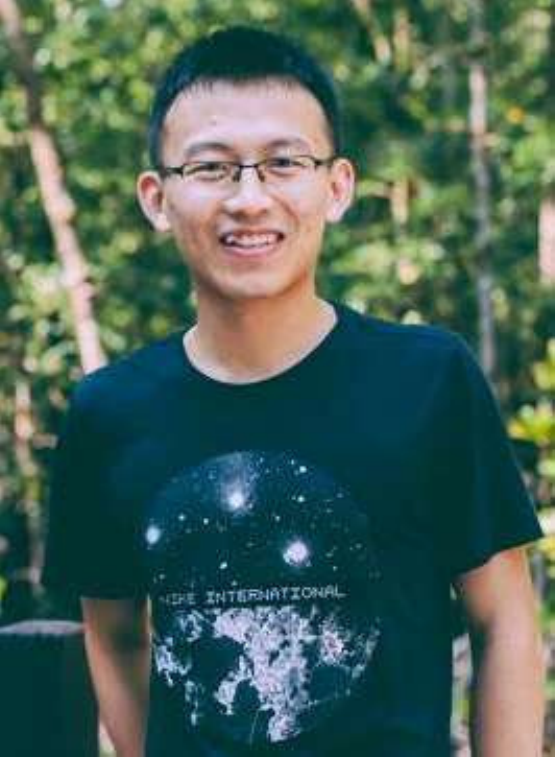}
\vspace{-0.5cm}
\end{figure}
\noindent{\bf Fuli Feng} is a Professor in University of Science and Technology of China. He received Ph.D. in Computer Science from National University of Singapore in 2019. His research interests include information retrieval, data mining, causal inference and multi-media processing. He has over 60 publications appeared in several top conferences such as SIGIR, WWW, and SIGKDD, and journals including TKDE and TOIS. He has received the Best Paper Honourable Mention of SIGIR 2021 and Best Poster Award of WWW 2018. Moreover, he has been served as the PC member for several top conferences including SIGIR, WWW, SIGKDD, NeurIPS, ICML, ICLR, ACL and invited reviewer for prestigious journals such as TOIS, TKDE, TNNLS, TPAMI. \\
E-mail: fulifeng93@gmail.com

\end{document}